\documentclass[12pt]{article}
\usepackage
[left=2.40cm, right=2.40cm, top=2.40cm, bottom=2.40cm]
{geometry}
\usepackage{amsmath,amssymb,amsfonts,amsxtra,mathrsfs,graphics,graphicx,amsthm,epsfig,ytableau,bm,longtable,float,color,tikz,mathtools,xfrac,footnote,rotating,lscape}

\usepackage[debug,pageanchor=false]{hyperref}
\definecolor{link}{rgb}{0,0,1}
\hypersetup{colorlinks=true,linkcolor=link,citecolor=link,urlcolor=link,linktocpage}

\usepackage[titles]{tocloft}

\restylefloat{table}
\usepackage{dsfont}
\usepackage{multicol}
\usepackage{lipsum}
\usepackage{textgreek}
\usetikzlibrary{decorations.pathmorphing}
\usetikzlibrary{decorations.markings}
\usetikzlibrary{quotes,arrows.meta}
\usetikzlibrary{arrows, decorations.markings, calc, fadings, decorations.pathreplacing, patterns, decorations.pathmorphing, positioning}
\usepackage{tikz-cd}

\usepackage{fixmath} 
\usepackage{scalerel}
\newlength\bshft
\def\fakebold#1{\ThisStyle{\ooalign{$\SavedStyle#1$\cr%
  \kern-\bshft$\SavedStyle#1$\cr%
  \kern\bshft$\SavedStyle#1$}}}

\usetikzlibrary{positioning,shapes}
\usetikzlibrary{chains}
\usetikzlibrary{arrows,fit,decorations.pathreplacing}
\tikzstyle{every picture}+=[remember picture]
\tikzstyle{na} = [baseline=-.5ex]

\usepackage{empheq}
\usepackage{cite}
\usepackage{multirow}
\usepackage{booktabs}
\usepackage[american]{babel}

\usepackage[latin1]{inputenc}

\usepackage{array,booktabs}

\makeatletter
\newcommand{\vast}{\bBigg@{1}}
\newcommand{\Vast}{\bBigg@{5}}
\makeatother

\numberwithin{equation}{section}

\makeatletter
\@addtoreset{equation}{section}
\makeatother

\newcommand{\eg}{\textit{e.g.}}

\newcommand{\ie}{\textit{i.e.}}

\newcommand{\etc}{\textit{etc}}

\newcommand{\ii}{\mathrm{i}}
\newcommand{\?}{\;\!}

\newcommand{\be}{\begin{equation}} \newcommand{\ee}{\end{equation}}
\newcommand{\bea}{\begin{equation} \begin{aligned}} \newcommand{\eea}{\end{aligned} \end{equation}}

\newcommand{\Iprod}[2]{\langle {#1}, {#2} \rangle}

\def\U{\mathrm{U}}
\def\SO{\mathrm{SO}}

\def\SU{\mathrm{SU}}

\newcommand{\ex}{{\mathrm{e}}}
\newcommand{\rd}{\mathrm{d}}

\newcommand{\vol}{\mathrm{vol}}

\DeclareMathOperator{\Tr}{Tr}

\DeclareMathOperator{\re}{\mathbb{R}e}
\DeclareMathOperator{\im}{\mathbb{I}m}

\newcommand{\pd}{\partial}

\newcommand{\cA}{\mathcal{A}}
\newcommand{\cB}{\mathcal{B}}
\newcommand{\cC}{\mathcal{C}}

\newcommand{\cF}{\mathcal{F}}

\newcommand{\cH}{\mathcal{H}}
\newcommand{\cI}{\mathcal{I}}
\newcommand{\cJ}{\mathcal{J}}
\newcommand{\cK}{\mathcal{K}}
\newcommand{\cL}{\mathcal{L}}
\newcommand{\cM}{\mathcal{M}}
\newcommand{\cN}{\mathcal{N}}

\newcommand{\cR}{\mathcal{R}}

\newcommand{\bC}{\mathbb{C}}

\newcommand{\bP}{\mathbb{P}}

\newcommand{\bR}{\mathbb{R}}
\newcommand{\bW}{\mathbb{W}}
\newcommand{\bZ}{\mathbb{Z}}

\newcommand{\fp}{\mathfrak{p}}

\usepackage{bbold}

\usepackage[Symbol]{upgreek}
\usepackage{bm}

\usepackage{calligra}
\DeclareMathAlphabet{\mathcalligra}{T1}{calligra}{m}{n}

\makeatletter
\g@addto@macro\bfseries{\boldmath}
\makeatother

\makeatletter
\newcommand*{\rom}[1]{\expandafter\@slowromancap\romannumeral #1@}
\makeatother

\begin{document}

%
%
	\begin{titlepage}
	\
	\vspace{2cm}
		\begin{center}

		       	{\Large \bf{Rotating multi-charge spindles and their microstates}}
				\bigskip
				
					Seyed Morteza Hosseini,$^a$ Kiril Hristov$^{b,c}$ and Alberto Zaffaroni$^{d,e}$\\
				\bigskip

{\footnotesize
{\it
$^a$Kavli IPMU (WPI), UTIAS, The University of Tokyo, Kashiwa, Chiba 277-8583, Japan \\
$^b$Faculty of Physics, Sofia University, 5 James Bourchier Blvd., Sofia 1164, Bulgaria \\
$^c$INRNE, Bulgarian Academy of Sciences, 72 Tsarigradsko Chaussee, Sofia 1784, Bulgaria \\
$^d$Dipartimento di Fisica, Universit\`a di Milano-Bicocca, I-20126 Milano, Italy \\
$^e$INFN, sezione di Milano-Bicocca, I-20126 Milano, Italy
}
}

       \vskip .2in
       
{\footnotesize
       {\it E-mail:} \href{morteza.hosseini@ipmu.jp}{\texttt{morteza.hosseini@ipmu.jp}}, \href{khristov@phys.uni-sofia.bg}{\texttt{khristov@phys.uni-sofia.bg}}, \href{alberto.zaffaroni@mib.infn.it}{\texttt{alberto.zaffaroni@mib.infn.it}}
}       
       \vskip .5in
       {\bf Abstract } 
       \vskip .1in
       \end{center}

Some AdS$_3 \times M_7$ type IIB vacua have been recently proposed to arise from D3-branes wrapped on a spindle, a sphere with conical singularities at the poles. We explicitly construct a generalization of these solutions corresponding to a class of electrically charged and rotating supersymmetric black strings in AdS$_5 \times S^5$ with general magnetic fluxes on the spindle. We then perform a counting of their microstates using the charged Cardy formula. To this purpose, we derive the general form of the anomaly polynomial of the dual $\mathcal{N} = (0 , 2)$ CFT in two dimensions and we show that it can be obtained via a simple gluing procedure.

       \end{titlepage}

%
%

\setcounter{tocdepth}{2}

\hrule

\tableofcontents

\vspace{0.65cm}
\hrule

%
%

\section{Introduction}

The relation of the Cardy formula \cite{Cardy:1986ie} to the physics of black holes and black strings is an old subject. In this paper we consider near-horizon solutions of rotating black strings that can be embedded in AdS$_5\times S^5$. In five-dimensional language, these are expected to arise as supersymmetric domain walls  that interpolate between AdS$_5$ and a near-horizon region consisting of a warped  fibration of the Ba\~nados-Teitelboim-Zanelli (BTZ) metric over a two-dimensional compact space. By  adding momentum along the string direction and compactifying it on a circle we can obtain a black hole with a smooth near-horizon geometry in four dimensions. We recently constructed various examples of spherical string solutions that can be embedded in AdS$_5\times S^5$ (or AdS$_7\times S^4$) and successfully matched the entropy of the corresponding four-dimensional black holes with a microscopic counting based on the charged Cardy formula \cite{Hosseini:2019lkt,Hosseini:2020vgl}. We considered rotating and charged generalizations of well-known AdS$_3$ vacua in type IIB and M-theory obtained by compactifying D3- and M5-branes on a sphere with a topological twist  \cite{Maldacena:2000mw,Benini:2012cz,Benini:2013cda,Benini:2015bwz}.

In this paper we extend our analysis to another class of supersymmetric AdS$_3$ vacua in type IIB string theory which are based on D3-branes wrapped on a spindle, \ie\,the one-dimensional complex weighted projective space, $\bW \bP^1_{[n_1, n_2]}$, which is an orbifold with spherical topology and two conical singularities at the poles \cite{Gauntlett:2006af,Gauntlett:2006ns,Kunduri:2006uh,Kunduri:2007qy,Ferrero:2020laf}. An example of this class of solutions was originally found in \cite{Gauntlett:2006af} and later generalized in \cite{Gauntlett:2006ns} working in ten dimensions.  It can be seen as a ten-dimensional uplift of a AdS$_3\times \bW \bP^1_{[n_1, n_2]}$ supersymmetric solution of minimal gauged supergravity found in \cite{Kunduri:2006uh} and generalized in \cite{Kunduri:2007qy} to the case of multiple charges. Supersymmetry is realized  through a magnetic flux along the spindle but the theory is \emph{not} topologically twisted \cite{Ferrero:2020laf}.  Notably, the full ten-dimensional metric is completely regular, \ie\,the apparent conical singularities in five dimensions are smoothed out by the embedding of the spindle inside a seven-dimensional geometry \cite{Kunduri:2007qy,Ferrero:2020laf}. The authors of  \cite{Kunduri:2006uh,Kunduri:2007qy} described these solutions as potential horizons of ``unbalanced spherical black rings''. More recently the authors of \cite{Ferrero:2020laf}  successfully pursued the interpretation in terms of $\cN=4$ super Yang-Mills (SYM) theory wrapped on the spindle which gives rise to a two-dimensional  $\cN = (0, 2)$ conformal field theory (CFT). Quite remarkably, despite the presence of singularities,  they were able to match  the central charge computed holographically with the result obtained from integrating the anomaly polynomial of $\cN=4$ SYM on $\bW \bP^1_{[n_1, n_2]}$. 

In the first part of the paper, we generalize these solutions by adding rotation and general electric and magnetic charges. To this purpose, we work in a four-dimensional setting using the recent general construction of supersymmetric dyonic  rotating black holes in four-dimensional $\cN=2$ gauged supergravity
with vector multiplets \cite{Hristov:2018spe,Hristov:2019mqp}. Unlike the solutions with a topological twist  \cite{Maldacena:2000mw,Benini:2012cz,Benini:2013cda,Benini:2015bwz} that arise from the twisted branch discussed in \cite{Hristov:2018spe}, the spindle black holes arise from the untwisted branch \cite{Hristov:2019mqp}, which also contains the Kerr-Newman black holes in AdS$_4$. These kinds of solutions are usually discarded because of the conical singularities. Once these are allowed, many new solutions naturally appear. In particular, we construct the near-horizon of a  family of  dyonic rotating black spindles, depending on two independent magnetic fluxes, three electric charges and one angular momentum subject to a constraint.  We discuss in detail the conserved electromagnetic charges and angular momentum, as well as the Bekenstein-Hawking entropy of the solutions. Then we uplift them to a five-dimensional gauged supergravity, which is a consistent truncation of type IIB on AdS$_5\times S^5$, and, finally, to type IIB. 

In the last part of the paper, we successfully match the entropy with a microscopic counting of states using the charged Cardy formula, 
\be
 \label{eq:susy:Charged-Cardy0}
 \log \rho_\text{susy} (n_l,   J_A) \approx 2 \pi \sqrt{\frac{c_l}{6} \left( n_l - \frac{c_l}{24} - \frac 12 (k^{-1})^{AB} J_A J_B \right)} \, ,
\ee
for the density of states with energy $n_l$ and charges $J_A$ of a two-dimensional CFT. To use this formula we  need to compute the central charge $c_l$ and the levels $k_{AB}$ of the currents in the  CFT at large $N$. These  can be extracted from the two-dimensional anomaly polynomial, which, in turn, can be obtained by integrating the four-dimensional anomaly polynomial of $\cN=4$ SYM on the compactification manifold. As for twisted compactifications, we need to  include a background for the internal $\U(1)$ isometry, which becomes a global symmetry in the two-dimensional theory \cite{Bah:2019rgq,Bah:2019vmq,Hosseini:2020vgl}. The two-dimensional anomaly polynomial in the case of the static spindle with a single magnetic flux was derived  in \cite{Ferrero:2020laf}. Here, we generalize the derivation to the case of general magnetic charges and of general background fields for the flavor charges. As in \cite{Ferrero:2020laf}, the exact two-dimensional $R$-symmetry mixes with the rotational symmetry of the spindle. Interestingly, also in this case, despite the conical singularities, all the anomaly coefficients of the two-dimensional CFT  can be extracted from purely four-dimensional physics. For future reference, we also give the expression of the two-dimensional anomaly polynomial for a generic $\cN=1$ CFT. 

It is interesting to observe that the two-dimensional anomaly polynomial in the large $N$ limit can be obtained from the gluing formula \eqref{A2d:gluingG}. This has a clear counterpart in the  proposal for writing entropy functionals for AdS black holes and black strings by gluing gravitational blocks \cite{Hosseini:2019iad} and, given the analogy with holomorphic blocks in quantum field theory (QFT)  \cite{Beem:2012mb}, it could shed further light on the physics 
of these systems.

The paper is organized as follows. In section \ref{sec:spindlehorizonsfrom4d}, we discuss how to obtain supersymmetric spindle solutions from four-dimensional gauged supergravity and set the conventions for the rest of this work. In section \ref{sect:spindle:stu}, we construct the near-horizon of the general family of  dyonic rotating black spindles in the $stu$ model. We will focus explicitly on two examples, which will be useful later.  The first is a rotating, electrically charged  generalization of the static spindle discussed in   \cite{Ferrero:2020laf}. The second is a static spindle depending on general magnetic charges under the $\U(1)^3$ isometry of $S^5$, which was already considered in \cite{Ferrero:2020laf}. In section \ref{sec:chargedCardy}, we derive the two-dimensional anomaly polynomial of the corresponding $\cN = (0, 2)$ CFT and match  the Bekenstein-Hawking entropy
of the solutions previously found with a microscopic counting based on the charged Cardy formula. In section   \ref{sec:N1theories}, we consider the anomaly polynomial for the compactification of an $\cN=1$ CFT on the spindle and examine the case of universal spindle solutions of minimal and half-maximal supergravity that can be embedded in most AdS$_5$ compactifications with eight and sixteen supersymmetries, respectively.
We conclude with comments and discussion in section \ref{sect:Discussion and outlook}.

\section{5d spindle horizons from 4d supergravity} 
\label{sec:spindlehorizonsfrom4d}

In this section we outline our approach to finding supersymmetric near-horizon geometries in five-dimensional $\cN = 2$ gauged supergravity
that are most generally fibrations of the BTZ metric with the  spindle. The solutions that we present explicitly in the next section are generalizations of the direct product solutions of AdS$_3$
with the spindle with purely magnetic charges found in minimal gauged supergravity in \cite{Kunduri:2006uh} and in the gauged $stu$ model in \cite{Kunduri:2007qy}.
These solutions were found earlier from a ten-dimensional  perspective in \cite{Gauntlett:2006af,Gauntlett:2006ns}, but our approach leads to a more immediate comparison with \cite{Kunduri:2006uh,Kunduri:2007qy}.

Here, we present a way to generalize the known static near-horizon spindle solutions to include electric charges and rotation.
This is done using the connection between the five-dimensional supergravity models we consider and their corresponding four-dimensional reduction.
This step requires that we add momentum along the string direction and compactify it to a circle in order to arrive at a smooth near-horizon solution in four dimensions,
thus changing the global structure of AdS$_3$ to BTZ. It is precisely this step that allows us to add extra electric charges and angular momentum to these solutions, which have eluded the standard classifications of AdS$_3$ solutions in ten dimensions or direct five-dimensional searches.

We first emphasize the new features of the spindle metric that we focus on, before taking a more detailed look at the 4d/5d relation and the explicit 4d class of solutions found in \cite{Hristov:2019mqp} that we need here. The explicit results from this preliminary analysis are presented in the next section.

\subsection{The metric on the spindle}
\label{subsec:spindlemetric}

We set our goal to look for the so-called spindle horizons, and therefore need to first describe the characteristics that set apart these horizons when looking for a particular supergravity solution. As described in \cite{Ferrero:2020laf,Ferrero:2020twa}, the weighted projective space $\bW \bP^1_{[n_1, n_2]}$ is topologically a sphere with conical deficit angles $2 \pi (1-1/{n_{1,2}})$ at the poles with $n_1 \neq n_2$ coprime positive integers. The Euler characteristic of the spindle is then given by 
\be
	\label{spindle:Euler}
	\chi (\Sigma) = \frac1{4 \pi} \int_\Sigma R_\Sigma \vol_\Sigma = \frac1{n_1} + \frac1{n_2} \, ,
\ee
where for brevity we introduced the notation $\Sigma =  \bW \bP^1_{[n_1, n_2]}$. We should note that in the math literature this is considered a {\it bad} orbifold in the sense that it does not admit a manifold as a covering space. This also means that the spindle can never have a constant curvature metric, making the distinction with other horizon topologies more manifest.%
\footnote{Note however that rotating black holes with spherical horizons in four dimensions also generically do not admit a constant curvature metric, so the deficit angle criterion discussed below remains the ultimate distinction.}

Schematically, up to conformal rescaling and model-dependent constant factors,%
\footnote{See the next section for the explicit solutions and precise constants and details of the metric; they are not important for the general discussion here.}
the spindle part of the metrics that we are going to consider here has the form
\be
\label{eq:spindlemetric}
	{\rm d} s_\Sigma^2 \sim \frac{{\rm d} p^2}{P(p)} + \frac{P(p)}{p^2 + P(p)}\, {\rm d} z^2\, ,
\ee
where $z$ is a $\U(1)$ isometry of the full metric (the axis of rotation) and the function $P(p)$ is in the general form
\be
	P(p) = s \prod_{a=1}^{n} (p - p_a)\ , \qquad p_a < p_{a+1}\, .
\ee  
Here, $s$ is a choice of overall sign setting the asymptotics, and generically the order of the polynomial is $n=4$ for asymptotically AdS$_4$ solutions%
\footnote{This case is relevant for example for the accelerating and spinning AdS$_4$ black hole spindle in \cite{Ferrero:2020twa}.}  and $n=3$ for asymptotically AdS$_5$ solutions,
but various special cases can occur depending on the specific type of solutions, conserved charges, \etc.
The spindle, like the sphere, is a {\it compact space} and therefore the function $P(p)$ must be {\it positive semi-definite} and {\it bounded}. In practice this leaves very little room for choice: for example in the case $n = 2$ the only chance that we find a compact metric is when $s = -1$ and we restrict $p \in [p_1, p_2]$. Similarly for $n=4, s = +1$ we must choose $p \in [p_2, p_3]$. Let us now focus on the cubic case that is directly relevant for the spindle strings in five dimensions:
\be
	n = 3: \qquad  s = +1 \, , \, p \in [p_1, p_2]\, , \quad \text{ or } \quad s = -1\, , \, p \in [p_2,p_3] \, .
\ee
The two choices are physically equivalent, and with no loss of generality we are going to choose the former case, $s = +1$. Note that all the solutions we present below allow also for a {\it non-compact} hyperbolic metric with the choice $n=3, s = +1, p \in [p_3, \infty]$. We leave this potentially interesting parameter space of hyperbolic black strings for a future investigation.%
\footnote{Above we already imposed that the roots of $P(p)$ are real and non-degenerate since in the cubic case this is the only chance for obtaining a compact metric; in the degenerate case there exist even more exotic possibilities such as non-compact horizons with finite area \cite{Gnecchi:2013mja,Klemm:2014rda}.}

Let us now look at the resulting metric near the poles. Using the coordinate redefinition $R \equiv 2 \sqrt{p - p_l}$, we can write the behavior of the metric near the two roots $p = p_l$, $l=1,2$,
\be
	{\rm d} s_\Sigma^2 \sim \frac1{P'(p_l)} \left( {\rm d} R^2 + \frac{P'(p_l)^2}{4 p_l^2}\, R^2 {\rm d} z^2 \right) .
\ee
At each of the poles separately we can always choose the period $\Delta z$ such that we get the metric on the $\mathbb{R}^2$ in polar coordinates.
If this were possible simultaneously for a single choice of the period $\Delta z$ at the two poles we would find a smooth manifold, \ie\,the sphere.
In the cubic case however this is only possible if $p_1 = p_2$, which is already inconsistent with the initial requirement that the metric is compact.
Therefore, one always finds at least one conical deficit angle, prompting the initial name ``unbalanced ring'' in \cite{Kunduri:2006uh,Kunduri:2007qy}.
We can instead obtain the metric on the spindle with two orbifold points of the form $\mathbb{C}/\mathbb{Z}_{n_1}$ and $\mathbb{C}/\mathbb{Z}_{n_2}$ with two positive coprime integers if we impose that
\be
	\frac{P'(p_l)^2}{4 p_l^2} = \left(\frac{2 \pi}{n_l \Delta z}\right)^2  ,
\ee
at both poles, thus fixing the period $\Delta z$.
This identification is going to be vital for the uplift to ten dimensions where the conical singularities get smoothed out inside the bigger internal space.

Let us finish the present discussion by noting that in a sense the spindle horizons are a {\it generic} feature of BPS and thermal black hole/string solutions in AdS. The type of function $P(p)$ and conical deficit angles describing the spindle require no fine tuning of parameters and in this sense are expected to show up in various solutions. It is in fact the smooth spherical horizon that requires an extra condition to be met (namely that $n_1 = n_2$), and therefore the spherical case generally requires one additional constraint in the parameter space of solutions in comparison to the spindle.

\subsection{The 4d/5d relation}
\label{subsec:4d-5d}
We are interested in finding solutions to five-dimensional $\cN=2$ gauged supergravity coupled to $n_{\text{V}}$ abelian vector multiplets. The Lagrangian is determined from the cubic prepotential function
\be
 \label{5d:prepotential}
 \cF_{5\rd} (L^i) = \frac 1 6 c_{ijk} L^i L^j L^k = 1 \, ,
\ee
where the five-dimensional scalars $L^i$, $i=1,\ldots, n_{\text{V}}$, are \emph{real}
and $c_{ijk}$ is a fully symmetric tensor appearing in the Chern-Simons terms,
corresponding to the 't Hooft anomaly coefficients of the dual four-dimensional $\cN=1$ CFT \cite{Ferrara:1998ur,Benvenuti:2006xg}. The bosonic action is given by 
\bea
 S^{(5)} = \frac1{8 \pi G_\text{N}^{(5)}} \int_{\bR^{4,1}} & \bigg[ \frac12 R^{(5)} \star_5 1- \frac12 G_{i j} \rd L^i \wedge \star_5 \rd L^j - \frac12 G_{i j} F^i \wedge \star_5 F^j \\
 & - \frac{1}{12} c_{i j k} F^i \wedge F^j \wedge A^k + V \star_5 1 \bigg] \, ,
\eea
with $R^{(5)}$ being the Ricci scalar, $F^i \equiv \rd A^i$ is the five-dimensional Maxwell field strength, and $G_{i j}$ can be written in terms of $\cF_{5\rd}$,
\bea
 \label{metric:gauge:kinetic:V}
 G_{i j} = - \frac12 \partial_i \partial_j \log \cF_{5\rd} \Big|_{\cF_{5\rd} = 1} \, .
\eea
The scalar potential can be written in terms of the inverse metric on the scalar manifold as
\bea
 V(L) = 9 V_i V_j \left( L^i L^j - \frac12 G^{i j} \right) ,
\eea
where the constants $V_i$ denote the five-dimensional Fayet-Iliopoulos (FI) parameters that specify completely the model together with the tensor $c_{i j k}$.
For more details about the five-dimensional conventions see \cite{Looyestijn:2010pb}.

In the gauged $stu$ model ($n_{\text{V}} = 3$) that we are going to consider explicitly, the only nonzero intersection numbers are $c_{123}=1$ (and cyclic permutations). The FI parameters are given by
\be
	V_1 = V_2 = V_3 = \frac{g_{(5)}}{3}\, ,
\ee
where $g_{(5)}$ sets the length scale of AdS$_5$. This gauged $stu$ model admits an embedding in $\cN=8$ gauged supergravity and an uplift on the $S^5$ to ten dimensions \cite{Cvetic:1999xp}.
It also admits a truncation to the $st^2$ model ($L^2 = L^3$) that can be embedded in minimal $\cN=4$ supergravity and thus admits a larger set of string theory embeddings \cite{Gauntlett:2007sm,Cassani:2019vcl,Malek:2017njj}.
Finally, we could even go to the $t^3$ model ($L^1=L^2=L^3=1$), which is just minimal $\cN=2$ gauged supergravity admitting an uplift on any SE$_5$ and any other AdS$_5$ solution \cite{Gauntlett:2007ma}.  

Reducing the five-dimensional theory along the circle $x^5$ one obtains a four-dimensional $\cN = 2$ gauged supergravity theory
based on the prepotential
\be
 \label{4d:prepotential}
 \cF_{4\rd} (X^\Lambda) = \frac16 c_{ijk} \frac{X^1 X^2 X^3}{X^0} \, ,
\ee
where the four-dimensional scalars $X^\Lambda = (X^0 , X^i)$ are \emph{complex}\?\!.
The four-dimensional theory contains $n_{\text{V}}$ abelian vector multiplets,
parameterizing a special K\"ahler manifold $\cM$ with metric $g_{i \bar{j}}$,
besides the gravity multiplet.
The following are the rules for reducing the bosonic fields \cite{Andrianopoli:2004im,Gaiotto:2005gf,Behrndt:2005he,Cardoso:2007rg}:%
\footnote{In our conventions the four-dimensional and five-dimensional gauge fields are related by $A_{(5)}^i  = \sqrt{2} A_{(4)}^i$. We make this choice for convenience to land directly at the conventions used in \cite{Hristov:2019mqp} based on the four-dimensional BPS equations in \cite{Meessen:2012sr}.} 
\bea
 \label{4d:5d:relation}
 \rd s_{(5)}^2 & = \ex^{2 \varphi} \, \rd s_{(4)}^2 + \ex^{- 4 \varphi} \left( \rd x^5 - \sqrt{2} A_{(4)}^0 \right)^2 \, , \qquad && \rd x^5 = \rd w \, , \\
 A_{(5)}^i & = \sqrt{2} A_{(4)}^i + \re z^i \left( \rd x^5 - \sqrt{2} A_{(4)}^0 \right) \, , \\
 L^i & = \ex^{2 \varphi} \im z^i \, , && \ex^{-6 \varphi} = \frac16 c_{i j k}  \im z^i  \im z^j  \im z^k \, .
\eea
Here, $\varphi$ is the Kaluza-Klein (KK) scalar (called dilaton),
$\rd s_{(4)}^2$ is the four-dimensional line element,
$A_{(4)}^\Lambda$, $\Lambda=0,\ldots,n_{\text{V}}$, are the four-dimensional abelian gauge fields and $z^i = X^i/X^0$. The resulting four-dimensional bosonic action is then given by
\bea
 S^{(4)} =  \frac1{8 \pi G_{\text{N}}^{(4)}} \int_{\bR^{3,1}} & \bigg[ \frac{1}{2} R^{(4)} \star_4 1
 + \frac{1}{2} \im \cN_{\Lambda \Sigma} F^\Lambda \wedge \star_4 F^{\Sigma}
 + \frac{1}{2} \re \cN_{\Lambda \Sigma} F^{\Lambda} \wedge F^{\Sigma} \\
 & - g_{i \bar{ j}} D z^{i} \wedge \star_4 D \bar{z}^{\bar{j}}
 - V(z, \bar{z}) \star_4 1 \bigg] \, ,
\eea
where $V (z , \bar z)$ is the scalar potential of the theory,
\bea
 V(z, \bar{z}) = 
 g_\Lambda g_\Sigma \left( U^{\Lambda \Sigma} - 3 e^{\cK} \bar{X}^\Lambda X^\Sigma \right) ,
\eea
with $g_\Lambda$ being the four-dimensional constant FI parameters.
See more details about the other four-dimensional supergravity quantities appearing in the action in \eg\,\cite{Andrianopoli:1996cm}.
Just like in five dimensions, the four-dimensional action is completely specified by the choice of prepotential and FI parameters.
The five-dimensional gauged $stu$ model therefore gives rise to the cubic prepotential \eqref{4d:prepotential} with $c_{123}=1$ (and cyclic permutations). In the fermionic sector we do a direct Kaluza-Klein reduction (as opposed to Scherk-Schwarz, see \cite{Hosseini:2017mds,Benini:2020gjh}) and therefore find the following  four-dimensional FI parameters
\be
\label{eq:FIparamfrom5d}
	g_0 = 0\ , \qquad g_1 = g_2 = g_3 = g_{(4)}\, ,
\ee
where, in our conventions,
\be
	g_{(4)} = \sqrt{2}\, g_{(5)}\, .
\ee

We can therefore look for a spindle near-horizon geometry of the type AdS$_2$ fibered over $\Sigma$ satisfying the criteria from section \ref{subsec:spindlemetric}, which upon uplift becomes a fibration of BTZ over $\Sigma$.
This would constitute a near-horizon geometry of a spindle black string in the original five-dimensional $\cN = 2$ gauged $stu$ model specified above.
Notice, that we need the extra circle $w$ {\it not} to be fibered over the spindle, such that we keep the same spindle geometry in five dimensions.
We therefore impose
\be
\label{eq:nofibration}
	\int_\Sigma \rd A^0_{(4)} = 0\ ,
\ee
meaning that we need a vanishing magnetic charge $m^0 = 0$.
This is an additional requirement that we impose on the four-dimensional solutions we consider next.
Note that $m^0 \neq 0$ instead leads in general to a Lens space horizon in five dimensions, which will be explored elsewhere.

\subsection{BPS black holes in 4d: the untwisted branch}
\label{subsec:KNsol}

We have now effectively reduced our initial problem from five to four dimensions,
the upshot being that the general supersymmetric near-horizon solutions in four-dimensional $\cN=2$ gauged supergravity
with abelian vector multiplets have already been written down exhaustively in the presence of the most general set of electromagnetic charges and rotation. The BPS solutions we discuss here are derived by solving the first-order integrability conditions following from a timelike isometry \cite{Cacciatori:2008ek,Meessen:2012sr}, which were further rewritten in the formalism and conventions we follow here in \cite{Hristov:2018spe,Hristov:2019mqp}.
In particular, one finds two disjoint branches of solutions -- the twisted%
\footnote{The most general twisted near-horizon geometry was written down in \cite{Hristov:2018spe} generalizing \cite{Cacciatori:2009iz,Halmagyi:2014qza}.}
and the untwisted%
\footnote{The most general untwisted near-horizon geometry was written down in \cite{Hristov:2019mqp} generalizing the supersymmetric limit of the Kerr-Newman-AdS$_4$ black hole.}
solutions.
The twisted branch in the model described above was used in \cite{Hosseini:2019lkt} to generalize the static black strings of \cite{Benini:2013cda} to the case with electric charges and rotation.
It is straightforward to see that the requirement $m^0 = 0$ for the twisted branch leads to a quadratic form of the function $P(p)$ in \eqref{eq:spindlemetric} and eventually one finds that the twisted branch
can only lead to spherical and {\it not} to spindle horizons for five-dimensional black strings.
Therefore, we turn our attention completely to the untwisted, or Kerr-Newman-like branch of solutions that lead to a cubic polynomial for $P(p)$.
The situation here is in fact reversed and we only find spindle (as opposed to spherical) horizons with non-vanishing area.

Referring the reader for all technical details to the original reference, \cite{Hristov:2019mqp}, here we summarize the main features of the untwisted near-horizons needed for the presentation of the solutions. We take the following ansatz for the four-dimensional spacetime,
\be
 \rd s_4^2 = - e^{2 U} \big( \rd t + \omega \big)^2 + e^{- 2 U} \rd s^2_3 \, ,
\ee
with  a base metric 
\be\label{metric3d}
 \rd s_3^2 = e^{2 \sigma} \left( \frac{\rd p^2}{P(p)} + \frac{\rd r^2}{Q(r)} \right) + P(p) Q(r) \rd z^2 \, ,
\ee
where 
\be
e^{2 \sigma}= r^2 P(p) + p^2 Q(r)\, ,
\ee
and $\omega$ is the rotation one-form living on the base space. The physical scalars $z^i$ together with the scalar factor $e^U$ are more conveniently packaged in the symplectic notation 
\be
	\{ X^\Lambda; F_\Lambda\} = e^{-2 U} \cR + {\rm i} \cI\, , \qquad \cR = - \frac{1}{2 I_4 (\cI)} I_4' (\cI) = - \frac12 e^{4 U}  I_4' (\cI)\, ,
\ee
where $F_\Lambda \equiv \frac{\partial \cF_{4\rd} (X^\Lambda)}{\partial X^\Lambda}$ and in this parametrization the symplectic vector $\cI$ determines both the real and the imaginary part of the physical scalars.
In the second equality we also used the so-called quartic invariant, $I_4$, which is indispensable for solving the BPS equations. The equations to be solved are given in  \cite[(12)-(14)]{Hristov:2019mqp}.
More technical aspects of the formalism are summarized in \cite{Hristov:2018spe} and the appendices of \cite{Halmagyi:2014qza,Hosseini:2019iad}. In the $stu$ model
with cubic prepotential coming from the 5d reduction, \eqref{4d:prepotential}, the quartic invariant and all its derivatives are uniquely defined by the equality
\be
\label{eq:I4electricstu}
  I_4 (\{p^\Lambda; q_\Lambda \}) = 4\? q_0 p^1 p^2 p^3 - (p^i q_i)^2  + 2 \sum_{i<j}^3 q_i p^i q_j p^j
- p^0 \bigg( 4\? q_1 q_2 q_3 + p^0 (q_0)^2 + 2\? q_0\?  p^i q_i \bigg) \, .
\ee
Finally, the four-dimensional gauge fields $A$ (or $\{A^\Lambda ; A_\Lambda \}$, including both electric and magnetic gauge fields) can also be expressed in terms of the three-dimensional gauge fields $\cA = \{\cA^\Lambda ; \cA_\Lambda \}$ on the base space,
\be
 \label{full:F}
\{ F^\Lambda ; G_\Lambda \} \equiv \rd A = \rd \left( \zeta ( \rd t + \omega ) \right) + \cF = \rd \left( \zeta ( \rd t + \omega ) \right) + \rd \cA \, , \qquad \zeta = \cR - G\? p \? r \, .
\ee
Here, $G$ denotes the symplectic vector of gauging parameters defining the Lagrangian and obtained from the 5d reduction
\be
	G = \{ 0; 0 , g, g, g \} \, .
\ee
Note that here we omitted the subscript $(4)$ in comparison with \eqref{eq:FIparamfrom5d} for simplicity
and will consistently do so below as long as we stay in four dimensions.
It is important here to note that the specific cubic prepotential and gauging parameters lead to the identity
\be
	I_4 (G) = 0\, ,
\ee
which is always satisfied for models coming from a 5d reduction as it signifies that there cannot exist AdS$_4$ asymptotics in these models (instead there is a runaway hvLif$_4$ vacuum, see more in \cite{Hristov:2018spe}).

In the AdS$_2 \times \Sigma$ near-horizon geometry conformal invariance requires $Q(r)= R_0^2 r^2$, where $R_0$  is a free constant,
and
\be
 e^{2\sigma}= r^2 e^{2 \sigma_0}\, , \qquad  e^{2U}= r^2 e^{2 U_0}\, , \qquad  \omega = \frac{1}{r} \omega_0 \, ,
\ee
where $\sigma_0, U_0$ and $\omega_0$ are functions of $p$.

Taking an ansatz for $\cI$ in terms of a symplectic vector $\cH$ polynomial in $p$, 
\be 
\cI = e^{-2 \sigma} \cH \, , \qquad \cH= r \left (\frac{\cC}{\Xi} + p \, \cC_1 + p^2 \, \cC_2 + p^3 \, \cC_3  \right) ,
\ee
with a free constant $\Xi$, the full set of BPS equations are solved in terms of the free constant symplectic vector $\cC$.
The symplectic vectors $\cC_{1,2,3}$ determining the scalars, metric functions and the three-dimensional gauge fields $\cA$ are then explicitly fixed by $\cC$,
the gauging vector $G$ and the quartic invariant $I_4$ using \cite[(22)]{Hristov:2019mqp}.
The rotation one-form reads
\be
	\omega_0 = - \frac{P(p) e^{-2 \sigma_0}}{2 \Xi}\, \Iprod{G}{I_4'(\cC)}\, , 
\ee
and thus vanishes in the special case that the symplectic product between the symplectic vectors $G$ and $I_4'(\cC)$ vanishes.
In the explicit examples below this indeed corresponds to the static spindle string solutions in \cite{Kunduri:2006uh,Kunduri:2007qy,Ferrero:2020laf}.

Finally, let us again focus on the form of the function $P(p)$, given by
\be
	P(p) = e^{-2 \sigma_0} - R_0^2 p^2 = \Xi^{-1} \left(1 + 2 \Iprod{G}{\cC} p + k_2 p^2 +\frac12 \Iprod{I_4'(G)}{I_4'(\cC)} p^3  \right) ,
\ee
where we defined $k_2 \equiv \Iprod{G}{\cC}^2 - \frac14 I_4(\cC, \cC, G, G) - \Xi R_0^2\, .$
Note that in general $P(p)$ also has a quartic power proportional to $I_4(G)$ that vanishes identically in the model of interest. Lead by an interest in smooth spherical horizons, \cite{Hristov:2019mqp} imposed in addition that $\Iprod{G}{\cC} = \Iprod{I_4'(G)}{I_4'(\cC)} = 0$, which is where our present analysis diverges. The only requirement that we impose on the a priori arbitrary constant vector $\cC$ is dictated by the condition of vanishing fibration of the additional circle with the spindle, \eqref{eq:nofibration}.
We therefore arrive at the most general solution in the gauged $stu$ model corresponding to spindle black strings,
\be
\label{eq:mostgenC}
	\cC = \left\{0, \frac1{2 g a_1}, \frac1{2 g a_2}, \frac1{2 g a_3}; b_0, b_1, b_2, b_3 \right\} ,
\ee
where $a_{1,2,3}$, $b_{0,1,2,3}$ are arbitrary constants and we inserted the factors of $g$ for later convenience.
The choice of $\cC$ in turn fixes all the conserved electromagnetic charges, rotation and entropy of the spindle horizons, which we evaluate explicitly in a couple of special examples.
Note also that we can rescale some of the parameters by a change of coordinates. In the following examples, we use this freedom to set $R_0$ and $\Xi$ to a convenient value without loss of generality on the physical observables and then analyze the metric on the spindle according to the considerations in section \ref{subsec:spindlemetric}.

\section{Spindle black strings in the $stu$ model}
\label{sect:spindle:stu}

So far we outlined a general procedure for finding spindle horizons via a passage through four-dimensional horizons.
We formally wrote down the most general solution in the gauged $stu$ model corresponding to a spindle black string horizon in \eqref{eq:mostgenC}, parametrized by a number of free constants.
We now turn to discuss the physics behind the solution in a couple of different limiting cases. 

First, we look at the equal magnetic charges case (setting $a_1=a_2=a_3$ in \eqref{eq:mostgenC}), allowing for arbitrary electric charges and rotation.
Then, we consider carefully the static case with different magnetic charges but vanishing electric charges (setting  $b_1 = b_2 = b_3 = 0$ in \eqref{eq:mostgenC}). This solution was
already discussed in \cite{Kunduri:2007qy}. In both cases we elaborate on the spindle metric and evaluate the conserved electromagnetic charges and angular momentum, as well as the Bekenstein-Hawking entropy.
In the first case, the full ten-dimensional regularity was already shown in \cite{Kunduri:2006uh,Ferrero:2020laf} for a general SE$_5$ uplift in agreement with \cite{Gauntlett:2006af}.
We discuss the ten-dimensional solution in the second case that only allows an $S^5$ uplift,
where we again see that the conical singularities are smoothed out \cite{Kunduri:2007qy}, as was seen also directly in ten dimensions \cite{Gauntlett:2006ns}. 

At the end of this section we also discuss how to obtain the resulting entropy from the procedure of gluing gravitational blocks generalizing \cite{Hosseini:2019iad} to spindle horizons,
which will turn out to mimic the way that the anomaly polynomial of the holographic dual CFT factorizes at the poles of the spindle.

\subsection{Rotating dyonic spindles with equal magnetic charges}
\label{subsec:rotspindle}

In this section we present a rotating and electrically charged generalization of the solution in \cite{Kunduri:2006uh,Ferrero:2020laf}. As described in the previous section, we first present the solution in four-dimensional language where it corresponds to the near-horizon of a rotating black hole with electric and magnetic charges. To have a real and positive entropy, we introduce a momentum along the compactifying circle which shows up in four dimensions as an electric charge $q_0$.  Similarly to \cite{Hosseini:2019lkt,Hosseini:2020vgl}, upon uplifting to five dimensions, the solution becomes a fibration of BTZ and the spindle. We allow for arbitrary electric charges $q_i$, $i=1,2,3$, but we restrict to equal magnetic charges, for simplicity.

Solving the BPS equations  \cite[(22)]{Hristov:2019mqp} with the following choice of the symplectic vector $\cC$,
\be
 \cC = \left\{ 0, \frac{1}{2 a g}, \frac{1}{2 a g}, \frac{1}{2 a g} ; b_0, b_1, b_2, b_3 \right\} ,
\ee
and a convenient choice for the parameters $R_0$ and $\Xi$,%
\footnote{Here, we set $R_0 = 1$ and $\Xi = \frac{27}{4 a^3}$.}
leads to a metric specified by the following functions 
\bea
Q (r) &= r^2   \, , \qquad \qquad  ~~ P(p) = \frac{4}{27} (p + a)^3 - p^2 \, , \\
e^{2 \sigma_0}  &= \frac{4}{27} ( p + a )^3   \, , \qquad \?   \omega_0  = - \frac{a}{27 g} \left( 1 - \frac{27}{4} \frac{p^2}{( p + a )^3} \right) \sum_{i=1}^3 b_i\, , \\
 e^{2 U_0} & =\frac{4 g^{\frac{3}{2}} (p + a)^3}{\sqrt{ 8 b_0 ( p + a )^3 - a g \Pi \left( 4 ( p + a )^3 - a p^2 \right) - 2 g (a p )^2 \sum_{i=1}^3 b_i^2 }} \, , \\
\eea
where we defined
\be
 \Pi = \sum_{i = 1}^3 b_i^2 - 2 ( b_1 b_2 + b_2 b_3 + b_1 b_3 ) \, .
\ee

The regularity of the metric can be directly checked following the procedure in section \ref{subsec:spindlemetric}.
We however decide to first change to the coordinates of \cite{Ferrero:2020laf} for a simpler comparison. Using the variable
\be
 p = 3 y - a  \, ,
\ee
we can write the polynomial $P$ as 
\be\label{PP}
  P(y) = 4 y^3 - (3 y-a)^2 = - \left(3 y + 2 y^{3/2} - a \right) \left(3 y - 2 y^{3/2} - a \right) \, .
\ee
Near the zeros of $P(y)$, we  have conical singularities and it is impossible to obtain a smooth metric. We can instead obtain a metric
with two orbifold singular points of the form $\mathbb{R}^{1,1} \times \bC/\bZ_{n_1}$ and $\bR^{1,1}  \times \bC/\bZ_{n_2}$, where $n_1$ and $n_2$ are two positive coprime integers. 
Consider the part of metric described by the coordinates $(y,z)$. Near a zero $y_l$ of $P(y)$,
using $R =2 \sqrt{y - y_l}$, we can write
\be
 \rd s^2_{\Sigma} \sim \frac{1}{P'(y_l)} \left( \rd R^2 + \frac{ P'(y_l)^2}{36\,  e^{2 \sigma_0(y_l)}} \? R^2 \rd z^2 \right) .
\ee
Choosing the period $\Delta z$ for $z$ such that
\be
 \label{5d:regularity}
  \frac{ P'(y_l)^2}{36\,  e^{2 \sigma_0(y_l)}} \equiv \left( \frac{2 \pi}{\Delta z \? n_l} \right)^2 \, , \quad \text{ for } \quad l=1,2 \, ,
\ee
we obtain a metric with deficit angles $2\pi/n_1$ and $2\pi/n_2$ at  $y_1$ and $y_2$, respectively.
Taking the zeros  such that $3 y_l - 2 y_l^{3/2} - a=0$ for $l=1,2$,  we obtain, by consistency,
\be
 a = \frac{(n_1-n_2)^2 (2 n_1+n_2)^2 (n_1+2 n_2)^2}{4 \left(n_1^2+n_2 n_1+n_2^2\right)^3} \, ,
 \qquad \Delta z = 4 \pi \? \frac{ n_1^2+n_2 n_1+n_2^2}{3 n_1 n_2 (n_1+n_2)} \, ,
\ee
and 
\be
 y_1 = \frac{(n_1-n_2)^2 (2 n_1+n_2)^2}{4 \left(n_1^2+n_2 n_1+n_2^2\right)^2} \, , \qquad
 y_2 = \frac{(n_1-n_2)^2 (n_1+2 n_2)^2}{4 \left(n_1^2+n_2 n_1+n_2^2\right)^2} \, .
\ee
With $y \in [y_1, y_2]$ we obtain a positive definite metric on the orbifold $\Sigma = \bW \bP^1_{[n_1, n_2]}$. Observe that $y_1 < y_2$ for $n_1 < n_2$.

The symplectic vector of gauge fields $\cA = \{0 , \cA^i ; \cA_\Lambda\}$  is given by
\bea
 \label{symp:A:dyonic}
 \cA^i & = \frac{1}{6 g} \frac{3 y - a}{y} \? \rd z \, , \qquad \text{for } \quad i = 1, 2, 3 \, ,\\
 \cA_0 & = (3 y-a) \left[ b_0 + \frac{g a (3 y-a)}{4 (3 y)^3} \left( a^2 ( b_1 + b_2 + b_3)^2 - 9 \Pi y^2 \right) \right] \rd z \, , \\
 \cA_i & = \frac{a (3 y-a)}{4 (3 y)^3} \left( 36 y^2 b_i - (9 y^2-a^2 ) ( b_1 + b_2 + b_3 ) \right) \rd z \, , \quad \text{ for } \quad i = 1, 2, 3 \, .
\eea
Notice that $\cA$ are just the components of the gauge fields on the base. The full expression for the gauge fields is given in \eqref{full:F} and contains components along $\rd t$ that are too long to be reported. 

We can now evaluate the conserved charges. The electromagnetic charges  read
\be
 \Gamma = \frac1{4 \pi} \int \cF = \left\{ m^\Lambda; q_\Lambda \right\}  ,
\ee
with
\bea\label{rotcharges}
 & m^0 = 0 \, , \qquad m^i = \frac1{6 g} \bigg( \frac{1}{n_1} -\frac{1}{n_2} \bigg) \, , \quad \text{ for } \quad i = 1, 2, 3 \, , \\
 & q_i = \frac{(n_1-n_2)^3 (2 n_1+n_2)^2 (n_1+2 n_2)^2}{12^2 n_1 n_2 \left(n_1^2+n_2 n_1+n_2^2\right)^4} \? \bigg((n_1-n_2)^2 \sum_{j = 1}^{3} b_j - 12 \left(n_1^2 + n_1 n_2 + n_2^2\right) b_i \bigg) .
\eea
There is also a non-zero  $q_0$ but its expression is not illustrative and we shall not present its explicit form here. 
Notice that the  $R$-symmetry gauge field 
\be
 A^R = \frac{g}{2} \sum_{i=1}^3 A^i
\ee  has a flux
\be
 \label{fluxSusy}
 \frac{1}{2 \pi} \int_\Sigma F^R = \frac{1}{2} \bigg( \frac{1}{n_1}-\frac{1}{n_2} \bigg) \, ,
\ee
along the spindle. Since the gauge field $A^R$ is the one appearing in the covariant derivative of the gravitino, we see that supersymmetry is realized with a mechanism analogous to the one discussed in \cite{Ferrero:2020laf}.
In particular, the components of $A^R$ along the spindle in the gauge \eqref{A:prop:flux} precisely coincide with those in \cite{Ferrero:2020laf}.
 
The angular momentum can be computed via the Komar integral associated to the spacelike Killing vector $\xi = \pd_\phi$, 
\be\label{angmomentum}
 \cJ =-\frac1{8 \pi} \int_\Sigma \left( \star_4\, \rd \xi + 2 (\xi \cdot A^\Lambda) \rd A_\Lambda \right) ,
\ee
where $\phi = 2\pi z/\Delta z$ is the angular coordinate on the spindle with period $ 2 \pi$. The above equation is the symplectically covariant generalization of the formula presented recently in \cite[Appendix E]{Ferrero:2020twa}.
As already noticed there, the angular momentum evaluated at the horizon is not gauge invariant and depends crucially on the choice of pure gauge that one can add to the electric gauge fields. We employ the gauge transformation $A^\Lambda \to A^\Lambda + A^\Lambda_G\?$,
\be
 A_G^0 = 0 \, , \qquad A_G^i = - \frac{1}{3 g}\, \rd z \, , \quad \text{ for } \quad i = 1, 2, 3 \, ,
\ee
so that at the poles we find
\be
 \label{A:prop:flux}
 A^i ( y_l ) = - \frac{1}{3 g n_l}\, \rd \phi  \, , \quad \text{ for } \quad i = 1, 2, 3\, , \quad \text{ and } \quad l = 1, 2 \, ,
\ee
as will be justified later.
The angular momentum $\cJ$ is then given by
\be
 \cJ = \frac{1}{72 g} \? \frac{ (n_1-n_2)^3 (n_1+n_2) (2 n_1+n_2)^2 (n_1+2 n_2)^2}{n_1^2 n_2^2 (n_1^2+n_2 n_1+n_2^2)^3} \sum_{i=1}^3 b_i \, .
\ee

It is also interesting to observe that the conserved charges obey the following constraint
\be\label{constrBSR}
 \frac{2}{3 g} \sum_{i = 1}^3 q_i + \mathring \varepsilon \cJ = 0 \, , \qquad
 \mathring \varepsilon = 3 \? \frac{n_1 n_2 (n_1+n_2)}{n_1^2+n_2 n_1+n_2^2} \, .
\ee
Constraints among charges are  common for supersymmetric AdS black holes and occur both for twisted and Kerr-Newman ones.

Finally, the Bekenstein-Hawking entropy reads
\bea\label{BHrot}
 S_{\text{BH}} & = \frac{\text{Area}}{4 G_{\text{N}}^{(4)}} \\
 & = \frac{3}{8 G_{\text{N}}^{(4)}} \left(\frac{3 g}{2}\right)^{-\frac3{2}} \sqrt{b_0 + a g \bigg( 2 ( b_1 b_2 + b_2 b_3 + b_3 b_1 ) - \frac{1}{2} \bigg( 1 + \frac{a}{27}  \bigg) (b_1 + b_2 + b_3)^2 \bigg)} \? (y_2 - y_1) \Delta z \\
 & = \frac{\pi}{g G^{(4)}_{\text{N}}} \sqrt{ \frac{( n_2 - n_1 )^3}{18 g n_1 n_2 ( n_1^2 + n_1 n_2 + n_2^2 )} \bigg( q_0 - \frac{1}{2} \sum_{A, B = 1}^2 ( q_A - q_3 ) (\texttt{k}^{-1})_{AB} ( q_B - q_3 ) - \frac{\cJ^2}{2 \texttt{k}} \bigg)} \, ,
\eea
where
\be
 \texttt{k}_{AB} = \frac{1}{6 g} \left( \frac{1}{n_1} - \frac{1}{n_2} \right)
 \begin{pmatrix}
 2 & 1 \\
 1 & 2
 \end{pmatrix} \, , \qquad
 \texttt{k} = - \frac{1}{(3 g)^3} \left( \frac{1}{n_1^3}-\frac{1}{n_2^3} \right) \, .
\ee

The solution can be uplifted to five dimensions using the formula \eqref{4d:5d:relation}. When $\cJ=0$ and $q_i=0 $, $ i=1,2,3$,  by rescaling the time coordinate, we obtain%
\footnote{Recall that in our conventions  $g_{(5)}=g_{(4)}/\sqrt{2}$ and we denoted $g_{(4)}\equiv g$. We also used the expression $\zeta = r \left\{ - \frac{1}{2 b_0},0,0,0 ; 0, g a, g a, g a \right\}$, valid  for zero angular momentum and zero electric charges, to evaluate the time component of the gauge field $A^0$.}
\bea\label{qq} 
 \rd s^2_5 & = \frac{1}{g_{(5)}^2} \left( \frac{4 y}{9} \rd s_{\text{BTZ}}^2 + \frac{y}{P(y)} \rd y^2 + \frac{P(y)}{(6 y)^2} \rd z^2 \right) ,
\eea
with the extremal BTZ metric   
\bea
 \label{ds2:BTZ}
 \rd s_{\text{BTZ}}^2 = \frac14 \left(  \frac{\rd r^2}{r^2} - r^2 \rd t^2\right) + r_+^2 \left( \rd w + \frac{ r}{2  r_+} \? \rd t \right)^2  \, ,
\eea
where $r_+ = \frac{1}{2} (3 g)^{3/2} \sqrt{b_0}$ is related to the four-dimensional electric charge $q_0$.
Since BTZ is metrically equivalent to AdS$_3$, we can replace  $\rd s^2_{\text{BTZ}}$ with $\rd s^2_{\text{AdS}_3}$ in \eqref{qq}
and recover the  solution found in \cite{Ferrero:2020laf}, corresponding to a two-dimensional CFT with central charge 
\be
 c^{\text{CFT}} = \frac{N^2}{3} \frac{( n_2 - n_1 )^3}{n_1 n_2 (n_1^2+n_1 n_2+n_2^2 )} \, .
\ee
In the solution \eqref{qq} there is a momentum along the BTZ circle, which is necessary to have a black hole  without naked singularities 
in four dimensions. From the field theory point of view, we are considering states  in the CFT$_2$  with energy proportional to $q_0$. 

For non-vanishing angular momentum and electric charges, the five-dimensional metric is more complicated and it is a non-trivial fibration of the spindle and the BTZ metric.
The physical interpretation is however simple. The solution  is dual to an ensemble of states of the CFT$_2$  with energy proportional to $q_0$ and conserved abelian charges $q_i$ and $\cJ$.
And, indeed, the entropy \eqref{BHrot} is  strongly reminiscent of the charged Cardy formula \eqref{eq:susy:Charged-Cardy0} that captures the density of states of a CFT$_2$. We
confirm this interpretation in section \ref{sec:chargedCardy}.

\subsection{Static spindles with different magnetic charges}
\label{subsec:staticspindle}

We now look more closely at the case of three different magnetic charges and vanishing electric charges $q_i = 0$ ($b_1 = b_2 = b_3 = 0$ in \eqref{eq:mostgenC}).
This is the four-dimensional version of the solution presented in \cite{Kunduri:2007qy} with extra momentum along the string direction.
We first write the solution in four dimensions and then uplift it to five dimensions, where it becomes a warped product of BTZ and the spindle, and,  later on,  to a smooth solution in ten dimensions. 

We parametrize the symplectic vector $\cC$ in the following way,
\be
 \cC = \left\{ 0, \frac1{2 g a_1}, \frac1{2 g a_2}, \frac1{2 g a_3} ; \frac{b_0}{(2 g)^3 a_1 a_2 a_3} , 0, 0 , 0 \right\} .
\ee
which leads to a static solution with%
\footnote{Here, we choose for convenience $R_0 = \frac{1}{\sqrt{\Xi}}$ and $\Xi = \frac{1}{a_1 a_2 a_3}$.}
\bea
 Q (r) &= a_1 a_2 a_3 r^2  \, , \qquad & P(p)&= \prod_{i = 1}^{3} ( p + a_i ) - p^2 a_1 a_2 a_3 \, , \\
e^{2 \sigma_0} & = \prod_{i = 1}^{3} ( p + a_i )   \, , \qquad &   e^{2 U_0}  &=   \frac12 (2 g)^3 e^{\sigma _0} \sqrt{\frac{a_1 a_2 a_3}{b_0}} \, .
\eea

To find a metric on the spindle, we follow the procedure in section \ref{subsec:spindlemetric}. 
The cubic polynomial $P(p)$ has three real positive roots which we denote by $p_l$, $l=1,2,3$.
Near $p_l$, the $(p, z)$ part of the metric behaves as 
\bea
 \rd s^2_{\Sigma} & \sim \frac{1}{P'(p_l)} \left( \frac{\rd p^2}{p - p_l} + \frac{a_1 a_2 a_3}{e^{2 \sigma_0(p_l)}} \? (p - p_l) P'(p_l)^2 \? \rd z^2 \right) .
\eea
Using $R =2 \sqrt{p - p_l}$, we obtain
\be
 \rd s^2_{\Sigma} \sim \frac{1}{P'(p_l)} \left( \rd R^2 + \frac{a_1 a_2 a_3 P'(p_l)^2}{4 e^{2 \sigma_0(p_l)}} \? R^2 \rd z^2 \right) .
\ee
Choosing the period $\Delta z$ for $z$ such that
\be
 \label{5d:regularity}
 \frac{a_1 a_2 a_3 P'(p_l)^2}{4 e^{2 \sigma_0(p_l)}} \equiv \left( \frac{2 \pi}{\Delta z \? n_l} \right)^2 \, , \quad \text{ for } \quad l=1,2 \, ,
\ee
where $(n_1, n_2)$ are arbitrary coprime positive integers with $n_1 < n_2$, we obtain conical singularities with deficit angles $2\pi/n_1$ and   $2\pi/n_2$ at $p_1$ and $p_2$, respectively.
The roots of the cubic polynomial $P(p)$ can then be written  as
\bea
 p_1 & = - \frac{1}{18 n_1 n_2 \? ( n_1 + n_2 )} \bigg( \cL + 2 ( n_1 - n_2 ) \sqrt{ \frac{2 \pi}{\Delta z}} \bigg) \bigg( \cL - 2 ( 2 n_1 + n_2 ) \sqrt{ \frac{2 \pi}{\Delta z}} \bigg) \, , \\
 p_2 & = - \frac{1}{18 n_1 n_2 \? ( n_1 + n_2 )} \bigg( \cL + 2 ( n_1 - n_2 ) \sqrt{ \frac{2 \pi}{\Delta z}} \bigg) \bigg( \cL + 2 ( n_1 + 2 n_2 ) \sqrt{ \frac{2 \pi}{\Delta z}} \bigg) \, , \\
 p_3 & = - \frac{1}{18 n_1 n_2 \? ( n_1 + n_2 )} \bigg( \cL - 2 ( 2 n_1 + n_2 ) \sqrt{ \frac{2 \pi}{\Delta z}} \bigg) \bigg( \cL + 2 ( n_1 + 2 n_2 ) \sqrt{ \frac{2 \pi}{\Delta z}} \bigg) \, .
\eea
Here, we also included the third root of $P(p)$ for completeness and, for the ease of notation, we defined
\be
 \cL = \sqrt{4 (n_1 ^2 + n_1 n_2+ n_2^2 ) \frac{2 \pi}{\Delta z} + 6 ( a_1 + a_2 + a_3 - a_1 a_2 a_3 ) n_1 n_2 (n_1 + n_2)} \, .
\ee
We choose $p \in [p_1, p_2]$ to obtain a positive definite metric on the orbifold $\Sigma = \bW \bP^1_{[n_1, n_2]}$. Observe that $p_1 < p_2$ for $n_1 < n_2$.

The symplectic vector of gauge fields on the base is given by
\be \label{backgf}
 \cA =  \left \{ 0,  \frac{a_1 a_2 a_3}{2 g}\frac{p}{p + a_1} \? \rd z\, , \frac{a_1 a_2 a_3}{2 g}\frac{p}{p + a_2} \? \rd z\, ,\?  \frac{a_1 a_2 a_3}{2 g}\frac{p}{p + a_3} \? \rd z\, ; \? \frac{b_0}{(2 g)^3} p\? \rd z \, , 0 \, , 0 \, ,0 \right \} ,
\ee
and the complete expression for the gauge fields is given  by \eqref{full:F} with
\be
 \zeta = r \left\{ - \frac{4 g^3 a_1 a_2 a_3}{b_0} , 0, 0 , 0 ; 0 , g a_1, g a_2 , g a_3 \right\}  .
\ee
The corresponding electromagnetic charges 
\be
 \Gamma = \frac1{4 \pi} \int_\Sigma \cF = \left\{ m^\Lambda; q_\Lambda \right\} ,
\ee
are given by
\bea
 m^0 & = 0 \, , \qquad && q_0 = \frac{b_0}{32 \pi g^3} \? ( p_2 - p_1 ) \Delta z \, , \\
 m^i & = \frac{a_1 a_2 a_3 ( p_2 - p_1)}{8 \pi g} \? \frac{a_i}{( p_1 + a_i ) ( p_2 + a_i )} \? \Delta z \, ,
 \qquad && q_i = 0 \, , \quad \text{ for } \quad i = 1, 2, 3 \, .
\eea
Note the following useful relations among the magnetic charges
\bea\label{magneticSTU}
 & \sum_{i = 1}^3 m^i = \frac{1}{2 g} \bigg( \frac{1}{n_1}-\frac{1}{n_2} \bigg) \, , \\
 & \prod_{i = 1}^3 m^i = p_3^2 \left(\frac{1}{8 \pi  g} ( p_2 - p_1 ) \Delta z \right)^3 \, , \\
 & m^1 m^2 + m^2 m^3 + m^1 m^3 = \frac{( p_1 - p_3 ) ( p_2 - p_3 ) + p_1 p_2 p_3^2}{p_1 p_2} \left(\frac{1}{8 \pi  g} ( p_2 - p_1 ) \Delta z \right)^2 \\
 & \qquad \qquad \qquad \qquad \qquad  = - \frac{1}{(2 g)^2} \? \left( \frac{1}{n_1 n_2} - \frac{a_1 a_2 a_3 }{4 \pi} \? \left ( \frac1{n_1} + \frac1{n_2}\right )\Delta z \right) .
\eea
The first equation in \eqref{magneticSTU} ensures again that the $R$-symmetry field $A^R=g\sum_{i=1}^3 A^i/2$ has a flux
\be\label{fluxSusy2}
 \frac{1}{2 \pi} \int_\Sigma F^R = \frac{1}{2} \bigg( \frac{1}{n_1}-\frac{1}{n_2} \bigg) \, ,
\ee
along the spindle, which is necessary to enforce supersymmetry \cite{Ferrero:2020laf}.
The electric charge can be also rewritten as
\be
 q_0 = \frac{b_0}{(2 g)^2}\frac{m^1 m^2 m^3}{ \frac{1}{(2 g)^2 n_1 n_2} + m^1 m^2 + m^2 m^3 + m^1 m^3} \, .
\ee

With this information we can evaluate the Bekenstein-Hawking entropy
\bea\label{BHmag}
 S_{\text{BH}} & = \frac{\text{Area}}{4 G_{\text{N}}^{(4)}} = \frac{\sqrt{b_0}}{2 ( 2 g)^3 G_{\text{N}}^{(4)}} \? ( p_2 - p_1 ) \Delta z \\
 & = \frac{\pi}{g G_{\text{N}}^{(4)}} \sqrt{\frac{m^1 m^2 m^3}{ \frac{1}{(2 g)^2 n_1 n_2} + m^1 m^2 + m^2 m^3 + m^1 m^3} \? q_0} \, .
\eea

The spindle black string, using the uplift formula \eqref{4d:5d:relation}, can be recast as 
\bea\label{qqq}
 \rd s_5^2 & = \frac1{4 g_{(5)}^2} \bigg( \frac{4 e^{\frac23 \sigma_0}}{a_1 a_2 a_3} \?
 \rd s^2_{\text{BTZ}} + \frac{e^{\frac23 \sigma_0}}{P(p)} \? \rd p^2 + \frac{a_1 a_2 a_3 P(p) }{e^{\frac43 \sigma_0}} \? \rd z^2 \bigg) \, ,\\
 A^i_{(5)} & = \rho^i(p) \? \rd z \equiv \frac{a_1 a_2 a_3}{2 g_{(5)}} \frac{p}{p + a_i} \? \rd z \, , \qquad
 L^i = \frac{e^{\frac23 \sigma_0}}{p + a_i} \, , \quad \text{ for } \quad i=1,2,3 \, ,
\eea
with the extremal BTZ metric given in \eqref{ds2:BTZ} and $r_+ = \frac{\sqrt{b_0}}{2 \sqrt{2}}$.
Note that the dilaton is given by
\be
 e^{2 \varphi} = 2 g_{(5)} \sqrt{\frac{2 a_1 a_2 a_3}{b_0}} \, e^{- \frac13 \sigma_0} \, .
\ee

Since the BTZ is locally equivalent to AdS$_3$, we can replace  $\rd s^2_{\text{BTZ}}$ with $\rd s^2_{\text{AdS}_3}$ in \eqref{qqq}
and find a generalization of  the  AdS$_3$ solution found in \cite{Ferrero:2020laf} depending on general magnetic charges and discussed in \cite{Gauntlett:2006ns,Kunduri:2007qy}. We compute the central charge of the corresponding CFT$_2$ and explicitly match 
the entropy \eqref{BHmag} with the charged Cardy formula in section \ref{sec:chargedCardy}. Our solution reduces to the one in  \cite{Ferrero:2020laf} for $a_1=a_2=a_3$.%
\footnote{Define $p=\frac 49 a^3_{{\rm here}} y -a_{{\rm here}}$ where $a_1=a_2=a_3\equiv a_{{\rm here}}$ and identify $a_{{\rm there}}=\frac{27}{4 a^2_{{\rm here}}}$ and $z_{\text{there}} = a^3_{\text{here}} \? z_{\text{here}}$.}

We can uplift the solution to  ten dimensions using \cite[(2.1)]{Cvetic:1999xp}
\be
 \label{10d:ansatz}
 \rd s^2_{10} = \sqrt{\Delta} \, \rd s^2_{5} + \frac1{ g_{(5)}^2 \sqrt{\Delta}} \sum_{i=1}^3 \frac{1}{L^i} \left( \rd \mu_i^2 + \mu_i^2 \left( \rd \varphi_i +  g_{(5)} A_{(5)}^i \right)^2 \right) ,
\ee
where $\Delta = \sum_{i=1}^{3} L^i \mu_i^2$, and
\be
 \mu_1 = \sin \theta \, , \qquad \mu_2 = \cos \theta \sin \psi \, , \qquad \mu_3 = \cos \theta \cos \psi \, ,
\ee
so that
\be
 \sum_{i=1}^3 \mu_i^2 = 1 \, , \qquad \rd s^2_{S^5} = \sum_{i=1}^3 \left( \rd \mu_i^2 + \mu_i^2 \rd \varphi_i^2 \right) .
\ee
The metric \eqref{10d:ansatz} can then also be rewritten as 
\bea
 \rd s_{10}^2 & = \frac{\sqrt{\Delta} \, e^{\frac23 \sigma_0}}{4 g_{(5)}^2}
 \left[ \frac{4}{a_1 a_2 a_3} \rd s^2_{\text{BTZ}}
 + \frac{\rd p^2}{P}
 + \frac{a_1 a_2 a_3 P}{e^{2 \sigma_0}} \bigg( 1 + \frac{\sqrt{2} g_{(5)} \? p \? e^{\frac23 \sigma_0}}{P \Delta} \sum _{a=1}^3 \rho^a \? \mu_a^2 \bigg) \text{D} z^2 \right] \\ &
 + \frac{1}{ g_{(5)}^2 \sqrt{\Delta}} \? \rd s^2_{B_5} \, ,
\eea
where
\bea
 \text{D} z & \equiv \rd z + \frac{2 \sqrt{2} \? p \? e^{\frac23 \sigma_0}}{\Delta ( P + e^{2 \sigma_0} )} \sum _{a=1}^3 \mu_a^2 \? \rd \varphi_a \, , \\
 \rd s^2_{B_5} & = \sum _{a=1}^3 \frac{1}{L^a} \left( \rd \mu_a^2 + \mu_a^2 \? \rd \varphi_a^2 \right)
 + \frac{P - e^{2 \sigma_0}}{\Delta (P + e^{2 \sigma_0})} \bigg(\sum _{a=1}^3 \mu_a^2 \? \rd \varphi_a \bigg)^2 \, ,
\eea
which allows to see the transverse seven-dimensional metric as a fibration of the angular variable $z$ over a six-dimensional base. 

In the ten-dimensional metric the $\U(1)^3$ torus in $S^5$ is non-trivially fibered over the spindle with Chern numbers determined by $m^i$. 
The regularity of the ten-dimensional solution has been explicitly checked  in \cite{Gauntlett:2006af,Ferrero:2020laf} for $a_1=a_2=a_3$ and in \cite{Gauntlett:2006ns,Kunduri:2007qy} for the general case. In the case of equal magnetic charges, 
the Reeb direction  $\psi=\phi_1+\phi_2+\phi_3$  is fibered over the spindle parameterized by $(p,z)$ with connection $2 A$, where \eqref{fluxSusy} holds. This gives a Lens space
$S^3/\mathbb{Z}_{(n_2-n_1)/3}$ fibration  over $\mathbb{P}^2$.%
\footnote{The total space of the fibration over the weighted projective space  $\bW \bP^1_{[n_1 ,n_2]}$ with Chern number $\frac{1}{2\pi} \int F= r/(n_1 n_2)$
is the Lens space $S^3/\mathbb{Z}_{r}$ where $\mathbb{Z}_{r}$ acts as $(z_1,z_2) \rightarrow ( e^{2\pi i n_1/r} z_1, e^{2\pi i n_2/r} z_2)$ on $S^3=\{ (z_1,z_2)\in \mathbb{C}_2 : |z_1|^2 +|z_2|^2=1\}$
-- see for example Appendix A in \cite{Ferrero:2020twa}. The extra factor of $3$ comes from the $6\pi$ periodicity of $\psi$.}
When $n_1$ and $n_2$ are relatively prime and $n_1-n_2$ is a multiple of $3$, the fibration is regular \cite{Gauntlett:2006af,Ferrero:2020laf}.
In the general case, the Reeb direction is still fibered on the spindle with the same Chern number, which is fixed by supersymmmetry. In addition, the flavor symmetry directions $\phi_1-\phi_3$ and $\phi_2-\phi_3$ are further fibered over the base.
It follows from the analysis done in \cite{Gauntlett:2006ns,Kunduri:2007qy} that the metric is still regular, for example, if $2 g n_1 n_2 m^i \in \mathbb{Z}$ \cite{Kunduri:2007qy}, which just imposes further quantization conditions  on the {\it flavor} magnetic charges $m_1-m_3$ and $m_2-m_3$. We leave the full analysis of the quantization conditions for the future.

\subsection{Entropy function from gravitational blocks}

In this section we show  that can  write an entropy functional for the spindle solution by gluing gravitational blocks, 
thus confirming the general prescription introduced in \cite{Hosseini:2019iad}.

It has been shown in \cite{Hosseini:2019iad} that all known entropy functionals for AdS$_4$  black holes and dimensional reduction of AdS$_5$ black strings can be  written as
\be\label{gluing}
 \cI ( p^\Lambda , \lambda^\Lambda , \varepsilon) \equiv \frac{\pi}{4 G_{\text{N}}^{(4)}} \bigg(\sum_{\sigma = 1}^2 \cB \big(X^\Lambda_{(\sigma)} , \varepsilon_{(\sigma)} \big) -2 \ii \lambda^\Lambda q_\Lambda - 2 \varepsilon \cJ \bigg) \, ,
\ee
where $\lambda^\Lambda$ and $\varepsilon$ are the chemical potentials conjugated to the electric charges $q_\Lambda$ and the angular momentum $\cJ$, respectively, and the  \emph{gravitational block}
\be
 \cB (X^\Lambda , \varepsilon) \equiv - \frac{\cF_{4\rd}(X^\Lambda)}{\varepsilon} \, ,
\ee
is constructed in terms of the prepotential $\cF_{4\rd}$ of the corresponding gauged supergravity. The functional $\cI$ must be extremized with respect to the chemical potentials $\lambda^\Lambda$ and $\varepsilon$ subject to a constraint  and the extremum value is the entropy of the black hole. The details of the  gluing functions $(X^\Lambda_{(\sigma)} , \varepsilon_{(\sigma)})$
and of the constraint  depend on the type of black hole,  either twisted or of Kerr-Newman type, and are discussed in \cite{Hosseini:2019iad}.
The construction is the gravitational dual of  the realization of three-dimensional supersymmetric partition functions obtained by gluing holomorphic blocks \cite{Beem:2012mb}, and fusing partition functions on hemispheres.
The two factors $\sigma=1$ and $\sigma=2$ correspond, in this language, to the North and South hemisphere of the horizon geometry $S^2$.

In the case at hand, the prepotential is given by
\be
 \cF_{4\rd} =\frac{X^1 X^2 X^3}{X^0} \, ,
\ee
and the sphere is replaced by the orbifold $\Sigma = \bW \bP^1_{[n_1, n_2]}$. The  gluing that works for the spindle is
\bea\label{Sgluing}
 X^0 & = \lambda^0 \, , \\
 X^{i}_{(1)} & = \lambda^i - \ii \? \varepsilon \? \bigg( m^\Lambda - \frac{1}{6 g} \chi \bigg) , \qquad && \varepsilon_{(1)} = \varepsilon \, , \\
 X^{i}_{(2)} & = \lambda^i + \ii \? \varepsilon \? \bigg( m^\Lambda+ \frac{1}{6 g} \chi \bigg) , && \varepsilon_{(2)} = - \varepsilon  \, ,
\eea
where $\chi$ is the Euler number of spindle, given in \eqref{spindle:Euler},
and the chemical potentials are constrained to satisfy
\be
 g_\Lambda \lambda^\Lambda = g \sum_{i=1}^3 \lambda^i =2 \, .
\ee
One can explicitly check that the entropy of both the rotating and magnetically charged spindle black holes, \eqref{BHrot} and \eqref{BHmag}, can be
obtained by extremizing the entropy functional \eqref{gluing}.

We can shed some light on the form of the gluing  \eqref{Sgluing} by considering first the case of equal magnetic charges $m^i = \frac1{6 g} \big( \frac{1}{n_1} -\frac{1}{n_2}\big)$ with $i=1,2,3$.  
The gluing of the components $i=1,2,3$ simplifies to
\bea\label{Sgluing2}
 X^{i}_{(1)} & = \lambda^i + \ii \? \varepsilon \? \frac{1}{3 g n_2}, \qquad && \varepsilon_{(1)} = \varepsilon \, , \\
 X^{i}_{(2)} & = \lambda^i + \ii \? \varepsilon \?  \frac{1}{3 g n_1}  , && \varepsilon_{(2)} = - \varepsilon \, .
\eea
We see that supersymmetry requires shifting $\lambda^i$ by a quantity proportional to the value of $R$-symmetry gauge field at the two hemispheres,   see  \eqref{A:prop:flux}. This is a natural generalization of what happens 
 for  twisted  black holes \cite{Hosseini:2019iad}, which motivated our ansatz. In the general case, we can write 
\bea
 \label{Sgluing22}
 X^{i}_{(1)} & = \lambda^i - \ii \? \varepsilon \? \bigg( s^i - \frac{1}{3 g n_2} \bigg) , \qquad && \varepsilon_{(1)} = \varepsilon \, , \\
 X^{i}_{(2)} & = \lambda^i + \ii \? \varepsilon \? \bigg( s^i + \frac{1}{3 g n_1}  \bigg) , && \varepsilon_{(2)} = - \varepsilon  \, ,
\eea
where, using \eqref{magneticSTU}, we parameterized 
\be
 m^i = s^i + \frac1{6 g} \bigg( \frac{1}{n_1} -\frac{1}{n_2} \bigg) \, , \quad \text{ for } \quad i = 1, 2, 3 \, ,
\ee
with $\sum_{i=1}^3 s^i = 0$. The $s^i$ can be seen as magnetic fluxes for the {\it flavor} symmetries of $\cN=4$ SYM. They enter in the gluing formula in analogy with the examples discussed in \cite{Hosseini:2019iad}.

The  form of the gluing \eqref{Sgluing} should follow from the gluing of holomorphic blocks in a field theory computation. The identification and evaluation of the supersymmetric index relevant for the spindle is left for future work.
It is not completely obvious what partition function we should consider in the presence of singularities.
We can however observe that the  form of the gluing  \eqref{Sgluing}  suggests that the relevant index is obtained by considering two hemisphere partition functions with 
the insertion of background magnetic fluxes $\frac{1}{3 g n_1}$ and $\frac{1}{3 g n_2}$, respectively, and gluing them together in the spirit of \cite{Beem:2012mb}.

We will see the  interpretation of the gluing in terms of the field theory anomaly polynomial in the next section.

\section{The charged Cardy formula and the spindle microstates}
\label{sec:chargedCardy}

In this section we provide a microscopic counting of the states of the black spindle using the charged  Cardy formula.

The black spindle horizon solutions discussed in this paper are dual to an ensemble of states of a $(0,2)$ supersymmetric CFT (SCFT) with  energy $L_0=n_l$ and charges $J_A$ under a set of abelian symmetries.
The density of supersymmetric states can be derived from the modular transformation of the elliptic genus and it is given by the charged Cardy formula  \cite{Hosseini:2020vgl}
\be
 \label{eq:susy:Charged-Cardy}
 \log \rho_\text{susy} (n_l,   J_A) \approx 2 \pi \sqrt{\frac{c_l}{6} \left( n_l - \frac{c_l}{24} - \frac 12 (k^{-1})^{AB} J_A J_B \right)} \, ,
\ee
where $c_l$ is the left-moving central charge and $k_{AB}$ is the matrix of levels of the abelian currents
\be
 \label{signconventions}
 k_{AB} = -  \Tr \gamma_3 J_A J_B \, .
\ee

All information needed to evaluated the Cardy formula can be extracted from the
two-dimensional anomaly polynomial $ \cA_{2\rd} $. In particular, the levels coincide (up to a sign)  with the 't Hooft anomaly coefficients, $ k_{AB} = - {\cal A}_{AB} $, defined as%
\footnote{As in \cite{Hosseini:2020vgl}, we use  notations where supersymmetry is realized in the anti-holomorphic sector
and the 2d chirality matrix $\gamma_3$ is taken to be positive on anti-holomorphic fermionic movers. We also choose the signs in such a way that the level matrix $k_{AB}$ in a unitary theory is positive definite for holomorphic currents.} 
\be
 \cA_{2\rd} = \frac 12 {\cal A}_{AB} c_1(F^A) c_1(F^B) + \ldots \, .
 \label{2dk}
\ee
 The exact $R$-symmetry of the two-dimensional CFT  is a linear combination of the abelian symmetries, $R=\sum  \mathring  \delta_A J_A$ that can be found by extremizing the trial central charge
\be c_r(\delta_A) = 3 \Tr \gamma_3R(\delta)^2 = 3 {\cal A}_{AB} \delta_A \delta_B \, ,\ee
where $R(\delta)=\sum \delta_A J_A$, with respect to the mixing parameters $\delta_A$. The restriction of the anomaly polynomial to the exact $R$-symmetry thus reads
\be
 \label{cc}
 \cA_{2\rd} = \frac{c_r}{6} c_1(R)^2 \, ,
\ee
where $c_r=c_r(\mathring \delta) \equiv c^{\text{CFT}}$ is the exact central charge. Since we work in the holographic regime, $c=c_l=c_r$.

We now compute the anomaly polynomial for the CFTs dual to the spindle solution and match the charged Cardy formula with the gravitational entropy.

\subsection[\texorpdfstring{$\cA_{2\rd}$}{A[2d]} for \texorpdfstring{$\cN = 4$}{N=4} super Yang-Mills on the spindle]{$\cA_{2\rd}$ for $\cN= 4$ super Yang-Mills on the spindle}
\label{2dan}

The $\cN = (0,2)$ SCFT is obtained by compactifying a set of $N$ D3-branes on the spindle,
and we expect to read off the 2d anomaly polynomial from the integration of the 4d anomaly polynomial of $\cN=4$ SYM on $\Sigma = \bW \bP^1_{[n_1 ,n_2]}$
in the presence of magnetic charges for the $\U(1)^3\subset \SO(6)$ global symmetries of the four-dimensional theory.
The two-dimensional theory  has an extra abelian symmetry, in addition to the $\U(1)^3$ inherited from the four-dimensional parent theory,
that arises from the $\U(1)$ isometry of the spindle.
The corresponding conserved charge is what we would call angular momentum from the higher-dimensional perspective.
As noticed in  \cite{Ferrero:2020laf}, the isometry along the spindle mixes in a non-trivial way with the $R$-symmetry of the two-dimensional theory
and we need to take it properly into account.
The inclusion of the symmetries coming from the internal geometry in the anomaly polynomial of the lower-dimensionsional CFT
has been extensively discussed in \cite{Bah:2019rgq,Bah:2019vmq,Hosseini:2020vgl} in the case of $S^2$.
The generalization to the spindle was discussed in \cite{Ferrero:2020laf}.
Here we need to further generalize it to the case of arbitrary charges.

Consider $\cN=4$ SYM   on $\Sigma$.
We use a basis of the $\U(1)^3\subset \SO(6)_R$ symmetry assigning charge $+1$ to chiral superfields $\Phi_{1,2,3}$, respectively;
we call their generators and field strengths as $Q_{1,2,3}$ and $F_{1,2,3}$, respectively.
The 4d anomaly polynomial in the large $N$ limit, for the gauge group $\SU(N)$, is
\be
 \cA_{4\rd} =  \frac{N^2}2 c_1(F_1)c_1(F_2)c_1(F_3) \, .
 \label{eq:4dN=4SYM}
\ee

The gravity solution corresponds to the situation where  we turn on background gauge fields $A_i = \rho_i(p) \rd \phi$ on the spindle with  magnetic fluxes
\be
 \frac{1}{2\pi} \int \rho^\prime_i (p) \rd p \? \rd \phi = \rho_i(p_2)-\rho_i(p_1) = \fp^i  \, , \quad \text{ for } \quad i = 1, 2, 3 \, ,
\ee
where, for convenience, we normalize the angle $\phi = 2\pi z/\Delta z$ to have period $ 2 \pi$.  As  in  \cite{Ferrero:2020laf}, the supersymmetry constraint \eqref{fluxSusy2} requires a flux $\frac 12\left ( \frac{1}{n_1}-\frac{1}{n_2}\right )$ for the $R$-symmetry which in our notations translate to\footnote{The field theory background fields $A_i$ are identified with $g_{(4)} A_{(4)}^i=g_{(5)} A_{(5)}^i$ on the gravity side.}
\be
 \label{sum:p}
 \sum_{i = 1}^3 \fp^i = \frac{1}{n_1} - \frac{1}{n_2} \, .
\ee 
We need to pay attention to the choice of gauge. By a gauge transformation, we can always add  additive constants to the  functions $\rho_i(p)$.  We  work in the gauge where
\be
 \label{gauge}
 \rho_i(p_2) = \frac 1 2 \Big( \fp^i -\frac{1}{3} \chi \Big) \, , \qquad \qquad  \rho_i(p_1) = \frac 1 2 \Big( \! -\fp^i -\frac{1}{3}\chi \Big) \, .
\ee
In particular, the   $R$-symmetry background field satisfies
\be\label{Rpoles}
 A_\phi^R(p_2)= -\frac{1}{2n_2}\, , \qquad A_\phi^R(p_1)=-\frac{1}{2n_1}\, .
\ee
As argued in  \cite{Ferrero:2020laf}, this choice of gauge ensures that the Killing spinors are independent of $\phi$.%
\footnote{This requirement only fixes the gauge for the $R$-symmetry. We could use a different gauge for the {\it flavor} symmetries, under which the Killing spinors are neutral. This ambiguity can be reabsorbed in a shift of the chemical potentials $\Delta_i$ in the trial function $c_r(\epsilon, \Delta_i)$ and leads to the same physical prediction for the central charge. Notice however that a change of gauge also leads to a redefinition of the charge associated with the internal isometry.  The same is true in gravity, see \eqref{angmomentum}.}  Notice that this is the same gauge used to compute $\cJ$  in section \ref{subsec:rotspindle} (see  \eqref{A:prop:flux}). 
The functions $\rho_i(p)$ can be read off from section \ref{subsec:staticspindle}, after normalization and a gauge transformation,  but their explicit forms will not be important in the following.

In order to compute the two-dimensional anomaly polynomial, we also turn on  background fields $A_R$ and $A_J$   probing the $R$-symmetry and the internal $\U(1)$ isometry, respectively,
\be  A_i = \Delta_i A_R + \rho_i(p) (\rd \phi + A_J) \, , \quad \text{ for } \quad i=1,2,3 \, ,\ee
with curvature 
\be
 F_i = \Delta_i F_R + \rho_i^\prime(p) \rd p (\rd \phi + A_J) + \rho_i(p) F_J \, , \quad \text{ for } \quad i = 1, 2, 3 \, ,
\ee
where now $A_R$ and $A_J$ are fields in the two-dimensional theory and we have embedded the 2d $\U(1)_R$ symmetry in the direction $\Delta_i Q_i$ with $\sum_{i=1}^3 \Delta_i=2$. 
We can now substitute this expression into \eqref{eq:4dN=4SYM} and integrate it over the spindle to obtain the two-dimensional anomaly polynomial
\be 
 \cA_{2\rd} = \int_\Sigma  \cA_{4\rd} \, ,
\ee
as a function of the background fields $A_R$ and $A_J$.
All integrals in $p$ can be explicitly done and we obtain
\bea \cA_{2\rd} =  \frac{N^2}2 & \Big [  \frac 12  c_{1} (F_R)^2 \sum_{i\ne j \ne k} \Delta_i \Delta_j [  \rho_k(p_2)-\rho_k(p_1)] \\ 
& +  \frac 12 c_{1} (F_R)c_{1} (J) \sum_{i\ne j \ne k} \Delta_i  [  \rho_j(p_2) \rho_k(p_2)- \rho_j(p_1) \rho_k(p_1)]\Big ]\\
&+ c_{1} (J)^2 [\rho_1(p_2) \rho_2(p_2) \rho_3(p_2) -  \rho_1(p_1) \rho_2(p_1) \rho_3(p_1)]  \, .\eea
It is interesting to observe that the previous expression can be recast as a sum over fixed points 
\bea 
 \cA_{2\rd} =  \frac{N^2}2 &
 \bigg[\frac{1}{c_1(J)} \prod_{i=1}^3 \left ( \Delta_i c_1(F_R)+ \rho_i(p_2) c_1(J)\right )
 + \frac{1}{(-c_1(J))} \prod_{i=1}^3 \left ( \Delta_i c_1(F_R)+ \rho_i(p_1) c_1(J)\right ) \bigg] \, ,
\eea
 of the internal $\U(1)$ isometry. 

Using \eqref{gauge}, the 2d anomaly polynomial can be compactly written as a gluing formula
\bea
 \label{A2d:gluing}
 \cA_{2\rd} = \frac{16}{27 c_1(J)} \left( a_{4\rd} \big( \Delta_i^{(1)} \big) - a_{4\rd} \big( \Delta_i^{(2)} \big) \right) ,
\eea
with
\be  \label{A2d:gluing2}
 \Delta_i^{(1)} = \Delta_i c_{1} (F_R) + \frac{c_1(J)}{2} \Big( \fp^i - \frac{1}{3} \chi \Big) \, , \qquad \Delta_i^{(2)} = \Delta_i c_{1} (F_R) - \frac{c_1(J)}{2} \Big( \fp^i + \frac{1}{3} \chi \Big) \, .
\ee
Note that \eqref{A2d:gluing} becomes a quadratic polynomial in $c_1(F_R)$ and $c_1 (J)$, after taking the sum over fixed points.
Here, $a_{4\rd} (\Delta_i)$ is the 4d trial central charge in the large $N$ limit, which for $\cN = 4$ SYM  reads
\be
 a_{4\rd} (\Delta_i) = \frac{27}{32} N^2 \Delta_1 \Delta_2 \Delta_3  \, .
\ee

The equations \eqref{A2d:gluing2} are the field theory counterparts of the gravitational gluing \eqref{Sgluing} for the components $i=1,2,3$.

\subsection{The case with one magnetic charge}\label{sec:1charge}

The anomaly polynomial of the two-dimensional theory obtained by compactifying $\cN = 4$ SYM on the spindle with equal magnetic fluxes 
\be 
 \fp^1=\fp^2=\fp^3=  \frac{1}{3}\left( \frac{1}{n_1} - \frac{1}{n_2} \right)  ,
\ee
is then given by 
\bea\label{Aequal}
 \cA_{2\rd} = \frac{2 a_{4\rd}}{27} & \Big( 9 ( \Delta_1 \Delta_2 + \Delta_1 \Delta_3 + \Delta_2 \Delta_3 ) \left( \frac{1}{n_1} - \frac{1}{n_2} \right) c_1 (F_R)^2 \\
 & + \left( \frac{1}{n_1^3} - \frac{1}{n_2^3} \right) c_1 (J)^2 - 6 \left( \frac{1}{n_1^2} - \frac{1}{n_2^2} \right) c_1 (F_R) c_1 (J) \Big) \, .
\eea
Here, $a_{4\rd} = \frac{N^2}{4}$ is the \emph{exact} central charge of $\cN = 4$ SYM in the large $N$ limit and the chemical potentials $\Delta_i$, $i=1,2,3,$ for the $\U(1)^3 \subset \SO(6)_R$ are constrained by
\be
 \label{sum:Delta=2}
 \sum_{i = 1}^3 \Delta_i = 2 \, .
\ee
This expression coincides with \cite[(30)]{Ferrero:2020laf} after setting $\Delta_1=\Delta_2=\Delta_3=\frac23$.

We see that $J$ mixes non-trivially with the $\U(1)^3$  symmetries of $\cN=4$ SYM. A convenient to way extract  the trial central charge is to allow a mixing $c_1 (J) = \epsilon \? c_1 (F_R)$ and then compute  $c_r(\epsilon, \Delta_i)= 6 \cA_{2\rd}/c_1 (F_R)^2 $:
\bea\label{cequal}
 c_{r} (\Delta_i, \epsilon) = \frac{4 a_{4\rd}}{9}  \Big[ 9 ( \Delta_1 \Delta_2 + \Delta_1 \Delta_3 + \Delta_2 \Delta_3 ) \left( \frac{1}{n_1} - \frac{1}{n_2} \right)  
  + \left( \frac{1}{n_1^3} - \frac{1}{n_2^3} \right) \epsilon^2 - 6 \left( \frac{1}{n_1^2} - \frac{1}{n_2^2} \right) \epsilon \Big] \, .
\eea
This has to be extremized  over the set of chemical potentials $(\epsilon, \Delta_i)$, under the constraint \eqref{sum:Delta=2}. We obtain the critical points
\be
 \mathring \epsilon = \frac{3 n_1 n_2 ( n_1 + n_2 )}{n_1^2 + n_1 n_2 + n_2^2} \, , \qquad \mathring \Delta_i = \frac23 \, , \quad \text{ for } \quad i = 1, 2, 3 \, .
\ee
We can then read off the \emph{exact} central charge of the two-dimensional CFT \cite{Ferrero:2020laf}
\be
 c^{\text{CFT}} = \frac{4 a_{4\rd}}{3} \frac{( n_2 - n_1 )^3}{n_1 n_2 (n_1^2+n_1 n_2+n_2^2 )} \, .
\ee

We can  define two independent flavor charges, $K_1=Q_1-Q_3$ and $K_2= Q_2- Q_3$, and  we can  extract their matrix level $k_{AB}$, $A,B = 1,2,$ as well as  the  level $k$ of the $\U(1)$ rotational symmetry 
using \eqref{2dk}
\be
 k_{AB} = \frac{2 a_{4\rd}}{3} \left( \frac{1}{n_1} - \frac{1}{n_2} \right)
 \begin{pmatrix}
 2 & 1 \\
 1 & 2
 \end{pmatrix} \, , \qquad
 k = - \frac{4 a_{4\rd}}{27} \left( \frac{1}{n_1^3}-\frac{1}{n_2^3} \right) .
\ee

We can now compare these results with the gravity prediction obtained in section \ref{subsec:rotspindle}.  The massless supergravity vector fields $A^i$, $i = 1, 2, 3,$  are associated with the Cartan subalgebra $\U(1)^3\subset \SO(6)$. In particular, \eqref{rotcharges} implies 
\be
 \fp^i = 2 g m^i = \frac{1}{3} \bigg( \frac{1}{n_1} - \frac{1}{n_2} \bigg) \, , \quad \text{ for } \quad i = 1, 2, 3 \, .
\ee
We also need the following relations among 5d and 4d quantities \cite{Hosseini:2020vgl}
\bea
 \label{dictionary}
 G^{(5)}_{\text{N}} & = 2 \pi G^{(4)}_{\text{N}} \, , \qquad J = \frac{1}{2 G_{\text{N}}^{(4)}} \cJ  \, , \qquad Q_0  = \frac{1}{2 \sqrt{2} \? G_{\text{N}}^{(4)}} q_0 \, , \\
 Q_i & = \frac{1}{2 \sqrt{2} \? g_{(5)} G_{\text{N}}^{(4)}} q_i \, , \quad \text{ for } \quad i = 1, 2, 3 \, .
\eea
Finally, we will use the well-known holographic relation for AdS$_5 \times S^5$
\be
 N^2 = \frac{\pi}{2 g_{(5)}^3 G_{\text{N}}^{(5)}} \, .
\ee

The constraint \eqref{constrBSR} can be now interpreted as the fact that the black spindle has  charge zero with respect to the exact $R$-symmetry 
\be
 \mathring{R} = \sum_{i=1}^3 \mathring \Delta_i Q_i + \mathring \epsilon J \, ,
\ee
of the CFT.
An analogous phenomenon was observed for the rotating black strings discussed in \cite{Hosseini:2020vgl}.

At this point, the entropy of the rotating black spindle \eqref{BHrot} can be written as%
\footnote{Here, we absorbed the vacuum energy in the definition of $Q_0 = n_l - \frac{c_l}{24}$.}
\be
 S_{\text{BH}}\equiv \log \rho_\text{susy} (Q_0,  Q_A, J) = 2 \pi  \sqrt{\frac{c^{\text{CFT}}}{6} \left( Q_0 - \frac{1}{2} \sum _{A, B=1}^2 (Q_A - Q_3) (k^{-1})_{AB}(Q_B - Q_3) - \frac{J^2}{2 k} \right)} \, ,
\ee
in complete agreement with the charged Cardy formula \eqref{eq:susy:Charged-Cardy}. 

The charged Cardy formula can be reformulated as an extremization problem. The standard derivation of the Cardy formula extracts the density of states from 
the high-temperature behavior of the CFT partition function that  is uniquely fixed by modular transformations \cite{Cardy:1986ie}. Analogously, 
the asymptotic density of supersymmetric states can be extracted from the asymptotic behavior of the elliptic genus of the CFT, which is in turn fully determined  by the 't Hooft anomalies of the theory. 
In our particular context, the density of states can be  obtained via extremizing \cite{Hosseini:2020vgl}
\be
 \label{index:CFT}
 \cI_{\text{CFT}} ( \epsilon, \Delta_i ) = \frac{\pi^2}{6 \beta} c_r(\Delta_i, \epsilon) + \beta Q_0 + \ii \pi \sum_{i=1}^3 \Delta_i Q_i + \ii \pi \epsilon J \, ,
\ee
with respect to $\beta, \epsilon, \Delta_{1,2,3}$, under the constraint \eqref{sum:Delta=2}. In this language, the constraint on charges \eqref{constrBSR} arises from
imposing the reality condition
\be
 \im \cI_{\text{CFT}} \Big|_{\text{crit.}} = \ii \pi \bigg( \sum_{i=1}^3 \mathring \Delta_i Q_i + \mathring \epsilon J \bigg) = 0 \, ,
\ee
and we obtain $\log \rho_{\text{susy}} = \re \cI_{\text{CFT}} \Big|_{\text{crit.}}$. We note that
the  functional  \eqref{index:CFT} is the field theory counterpart of the entropy functional \eqref{gluing} based on gravitational blocks. Using  \eqref{A2d:gluing} and \eqref{A2d:gluing2},
it is easy indeed to check that the two extremization problems can be mapped onto each other by identifying $\lambda^\Lambda$ with $(\beta, \Delta_i)$ (up to a suitable rescaling). 

\subsection{The case with arbitrary magnetic charges}\label{genm}

In the case of arbitrary magnetic charges, the anomaly polynomial is given by
\bea
 \label{A2d:general}
 \cA_{2\rd} = a_{4\rd} & \bigg( 2 (\Delta_1 \Delta_2 \fp^3 + \Delta_1 \Delta_3 \fp^2 + \Delta_2 \Delta_3 \fp^1) c_1 (F_R)^2 \\ &
 + \left( \frac{1}{18} \left(\frac{1}{n_1}+\frac{1}{n_2}\right) \left(\frac{1}{n_1^2}-\frac{1}{n_2^2}\right) + \frac1{2} \? \fp^1 \fp^2 \fp^3 \right) c_1 (J)^2 \\ &
 - \frac{1}{3} \left( \frac{1}{n_2} + \frac{1}{n_1} \right) ( (\Delta_1+\Delta_2) \fp^3 + (\Delta_1+\Delta_3) \fp^2 + (\Delta_2+\Delta_3) \fp^1) c_1 (F_R) c_1 (J) \bigg) \, ,
\eea
with $\fp^i$, $i=1,2,3,$ being the fluxes through the spindle for the $\U(1)^3 \subset \SO(6)_R$ symmetry, satisfying
\be
 \label{sum:p}
 \sum_{i = 1}^3 \fp^i = \frac{1}{n_1} - \frac{1}{n_2} \, .
\ee
By allowing a mixing $c_1 (J) =  \epsilon \? c_1 (F_R)$ and extremizing the trial central charge $c_r(\epsilon, \Delta_i)= 6 \cA_{2\rd}/c_1 (F_R)^2 $ over the set of chemical potentials $(\epsilon, \Delta_i)$, under the constraint \eqref{sum:Delta=2}, we obtain the critical points
\bea
 \mathring \epsilon & =  \left(\frac{1}{n_2}-\frac{1}{n_1}\right)^2
 \frac{n_1 n_2 (n_1 + n_2 )}{n_1 n_2 \left((\fp^1)^2+(\fp^2)^2+(\fp^3)^2\right) + ( n_1^2 + n_2^2) (\fp^2 \fp^3 + \fp^1 \fp^2 + \fp^1 \fp^3)} \, , \\
 \mathring \Delta_1 & = \frac{1}{18} \left( \frac{1}{n_2}-\frac{1}{n_1}\right)^2
 \frac{3 n_1 n_2  (3 n_2 \fp^1+2 ) - n_1^2  (3 n_2 \fp^1 (6 n_2 \fp^1+3)-3 ) + 3 n_2^2}
 {n_1 n_2 \left((\fp^1)^2+(\fp^2)^2+(\fp^3)^2\right) + ( n_1^2 + n_2^2) (\fp^2 \fp^3 + \fp^1 \fp^2 + \fp^1 \fp^3)} \, , \\
 \mathring \Delta_2 & = \frac{1}{18} \left( \frac{1}{n_2}-\frac{1}{n_1}\right)^2
 \frac{3 n_1 n_2 (3 n_2 \fp^2+2 ) - n_1^2 (3 n_2 \fp^2 (6 n_2 \fp^2 + 3 )-3 ) + 3 n_2^2}
 {n_1 n_2 \left((\fp^1)^2+(\fp^2)^2+(\fp^3)^2\right) + ( n_1^2 + n_2^2) (\fp^2 \fp^3 + \fp^1 \fp^2 + \fp^1 \fp^3)} \, .
\eea
Using \eqref{cc}, we can then read off the \emph{exact} central charge of the two-dimensional CFT
\be
 \label{cCFT:stu}
 c^{\text{CFT}} = \frac{12 a_{4\rd} \? \fp^1 \fp^2 \fp^3}{\frac{1}{ n_1 n_2}+\fp^1 \fp^2+\fp^3 \fp^2+\fp^1 \fp^3}\, .
\ee
Using \eqref{dictionary}, we can then write the entropy of the magnetically charged spindle \eqref{BHmag} as
\be
S_{\text{BH}}\equiv \log \rho_\text{susy} (Q_0) = 2 \pi  \sqrt{\frac{c^{\text{CFT}}}{6}  Q_0 } \, ,
\ee
in perfect agreement with the Cardy formula.

Notice, that for arbitrary magnetic charges, the flavor symmetries   $K_1=Q_1-Q_3$, $K_2= Q_2- Q_3$ and $K_3=J$ mix in a non-trivial way. For completeness,
we give the matrix level $k_{AB}$, $A,B = 1,2,3,$
\bea
 k_{1,1} & = 4 a_{4\rd} \? \fp^2 \, , \qquad k_{1,2} = 2 a_{4\rd}(\fp^1+\fp^2-\fp^3) \, , \qquad
 k_{1,3} = -\frac{1}{3} a_{4\rd} \left( \frac{1}{n_1} + \frac{1}{n_2} \right) (\fp^1-\fp^3) \, , \\
 k_{2,2} & =4 a_{4\rd} \? \fp^1 \, , \qquad k_{2,3} = - \frac{1}{3} a_{4\rd} \left( \frac{1}{n_1} + \frac{1}{n_2} \right) (\fp^2-\fp^3) \, , \\
 k_{3,3} & = - a_{4\rd} \? \left( \frac{1}{9} \left(\frac{1}{n_1}+\frac{1}{n_2}\right) \left(\frac{1}{n_1^2} - \frac{1}{n_2^2}\right) +  \? \fp^1 \fp^2 \fp^3 \right) .
\eea

\section[\texorpdfstring{$\cA_{2\rd}$}{A[2d]} for general \texorpdfstring{$\cN = 1$}{N=1} theories on the spindle]{$\cA_{2\rd}$ for general $\cN=1$ theories on the spindle}
\label{sec:N1theories}

The black string solution in \cite{Ferrero:2020laf}, being a solution to minimal gauged supergravity, can be embedded in all AdS$_5$ string and M-theory compactifications,\footnote{There are  restrictions due to the quantization conditions of the charges and the regularity of the uplift that need to be discussed on a case-by-case basis.}
in particular in all the type IIB vacua AdS$_5\times$ SE$_5$ associated with five-dimensional regular Sasaki-Einstein manifolds SE$_5$.  It would be interesting to find spindle black string solutions  depending on more general charges and angular momentum
in such compactifications. While finding the explicit solution is a difficult task, there is a simple quantum field theory prediction for the anomaly polynomial and the central charge of the corresponding CFT$_2$. In this section, we briefly discuss the form of the anomaly polynomial for general theories. We also discuss the case of solutions of mimimal and half-maximal gauged supergravity that can be embedded in infinitely many type II and M theory compactifications.

\subsection{Integrating the anomaly polynomial} 
Consider the compactification of a  four-dimensional  $\cN=1$ SCFT with $d$ abelian global symmetries dual to AdS$_5\times$ SE$_5$  on $\Sigma$.
In the large $N$ limit, the 4d anomaly polynomial can be written as
\be
 \cA_{4\rd} =  \frac16 \sum_{i,j,k=1}^d c_{ijk} \? c_1(F_i)c_1(F_j)c_1(F_k) \, ,
 \label{eq:4dCFT}
\ee
where $F_i$ is a basis of $R$-symmetries with generators $Q_i$ and $c_{ijk}$ are related to the 't Hooft anomaly coefficients.
In the toric case, there is a quite explicit description of 
the generators $Q_i$ on fields and the anomaly coefficients $c_{ijk}$ in terms of toric data \cite{Franco:2005rj,Feng:2005gw,Butti:2005vn,Benvenuti:2005ja,Benvenuti:2006xg}.\footnote{We follow the conventions of \cite{Hosseini:2019use}, where more details can be found.  In particular, $c_{ijk}=\frac{N^2}{2} |\det (v_i,v_j,v_k)|$ where $v_i$ are the integer vectors defining the toric fan. For $\cN=4$ SYM, $c_{123}=N^2/2$ and for the conifold $c_{123}=c_{124}=c_{234}=c_{134}=N^2/2$. Notice that, in our normalizations,  the $F_i$ assign charge $+1$ to the superpotential. \label{footnote8}}
We turn on background gauge fields $A_i = \rho_i(p) \rd \phi$ on the spindle with  magnetic fluxes $\fp^i$ satisfying
\be
 \label{sum:pgen}
 \sum_{i = 1}^d \fp^i = \frac{1}{n_1} - \frac{1}{n_2} \, ,
\ee 
and work in the gauge
\be
 \label{gauge2}
  \rho_i(p_2) = \frac 1 2 \Big( \fp^i -\frac{r_0^i}{2} \chi \Big)\, , \qquad \qquad \rho_i(p_1) = \frac 1 2 \Big( \! -\fp^i -\frac{r_0^i}{2}\chi \Big) \, , 
\ee
where $\sum_{i=1}^d r_0^i=2$, which fixes the values of $R$-symmetry background field at the poles as in formula \eqref{Rpoles}.

We also turn on  background fields $A_R$ and $A_J$  for the $R$-symmetry and the internal $\U(1)$ isometry,
with curvature 
\be
 F_i = \Delta_i F_R + \rho_i^\prime(p) \rd p (\rd \phi + A_J) + \rho_i(p) F_J \, , \quad \text{ for } \quad i = 1, 2, \dots , d \, ,
\ee
where we have embedded the 2d $\U(1)_R$ symmetry in the direction $\Delta_i Q_i$ with $\sum_{i=1}^d \Delta_i=2$. 

The two-dimensional anomaly polynomial reads
\be 
 \cA_{2\rd} = \int_\Sigma  \cA_{4\rd} \, .
\ee
Repeating the same arguments as  in section \ref{2dan}, it is easy to see that the anomaly polynomial can be compactly written again as  
a gluing formula
\bea
 \label{A2d:gluingG}
 \cA_{2\rd} = \frac{16}{27 c_1(J)} \left( a_{4\rd} \big( \Delta_i^{(1)} \big) - a_{4\rd} \big( \Delta_i^{(2)} \big) \right) ,
\eea
where
\be  \label{A2d:gluing2G}
\Delta_i^{(1)} = \Delta_i c_{1} (F_R) + \frac{c_1(J)}{2} \Big( \fp^i - \frac{r_0^i}{2} \chi \Big) \, , \qquad \Delta_i^{(2)} = \Delta_i c_{1} (F_R) - \frac{c_1(J)}{2} \Big( \fp^i + \frac{r_0^i}{2} \chi \Big) \, ,
\ee
and we have defined the   4d trial central charge in the large $N$ limit
\be\label{gen-a}
 a_{4\rd} (\Delta_i) = \frac{9}{32} \sum_{i,j,k = 1}^d c_{ijk} \Delta_i \Delta_j \Delta_k  \, .
\ee

More explicitly, we can write 
 \bea
  \cA_{2\rd} & =
  \frac{16}{27}  \sum_{i=1}^d \frac{\partial a_{4\rd} (\Delta_i)}{\partial \Delta_i} \fp^i \?  c_1 (F_R)^2 
  -  \frac{4 \chi}{27} \sum_{i,j=1}^d \frac{\partial^2 a_{4\rd} (\Delta_i)}{\partial \Delta_i\partial \Delta_j} \fp^i r_0^j  \?  c_1 (F_R) c_1(J) \\
 & +  \frac {2}{81} \sum_{i,j,k=1}^d \frac{\partial^3 a_{4\rd} (\Delta_i)}{\partial \Delta_i\partial \Delta_j\partial \Delta_k} \Big[\fp^i \fp^j \fp^k + \frac{3\chi^2}{4} \fp^i r_0^j r_0^k \Big]  \?  c_1(J)^2 \, .
 \eea
By allowing a mixing $c_1 (J) =  \epsilon \? c_1 (F_R)$ and extremizing the trial central charge $c_r(\epsilon, \Delta_i)= 6 \cA_{2\rd}/c_1 (F_R)^2 $  under the constraint $\sum_{i=1}^d \Delta_i=2$, we can determine the exact central charge of the CFT$_2$.
 
Let us consider, for example, the Klebanov-Witten theory \cite{Klebanov:1998hh}. 
The manifold in this case is $Y_5 = T^{1,1}$.
The quiver contains two $\SU(N)$ gauge groups with two bi-fundamental chiral fields $A_i$ in the representation $({\bf N},\overline{{\bf N}})$ and two bi-fundamental chiral fields $B_i$ in the representation $(\overline{{\bf N}},{\bf N})$ with a quartic superpotential
\be
 \label{ABJMsuper}
 W = \Tr \big( A_1B_1A_2B_2 - A_1B_2A_2B_1 \big) \, .
\ee
We introduce four chemical potentials $\Delta_I$ and fluxes $\fp^I$, one for each of the four fields $\{A_i,B_i\}$, associated with the four global symmetries of the theory and
satisfying
\be
\sum_{I=1}^4 \Delta_I = 2 \, , \qquad\qquad  \sum_{I = 1}^4 \fp^I = \frac{1}{n_1} - \frac{1}{n_2} \, .
\ee
The  4d trial central charge in the large $N$ limit reads%
\footnote{See footnote \ref{footnote8} for the 't Hooft anomaly coefficients.}
\be
 a_{4\rd} (\Delta_i) = \frac{27}{32} N^2 (\Delta_1 \Delta_2 \Delta_3 + \Delta_1 \Delta_4 \Delta_3 + \Delta_2 \Delta_4 \Delta_3 + \Delta_1 \Delta_2 \Delta_4 ) \, .
\ee
The exact  $R$-symmetry corresponds to $\Delta_I=1/2$ and the 
exact central charge is given by
\be
 a_{4\rd} = \frac{27}{64} N^2 \, .
\ee

Using $r_0^i= 1/2$, we can write the anomaly polynomial
\bea
 \frac{27}{a_{4\rd}} \? \cA_{2\rd} & = 32 \Big( \! \left(\Delta _3 \Delta _4 + \Delta _2 \Delta _3 + \Delta _2 \Delta _4  \right) \fp^1
 + \left(\Delta _1 \Delta _3 + \Delta _1 \Delta _4 + \Delta _3 \Delta _4 \right) \fp^2 \\
 & + \left(\Delta _2 \Delta _4 + \Delta _1 \Delta _2 + \Delta _1 \Delta _4 \right) \fp^3
 + \left(\Delta _1 \Delta _2 + \Delta _1 \Delta _3 + \Delta _2 \Delta _3 \right) \fp^4 \Big) c_1 (F_R)^2 \\
 & + \frac{1}{2} \left(3 \left(\frac{1}{n_1}-\frac{1}{n_2}\right) \left(\frac{1}{n_1} + \frac{1}{n_2}\right)^2 +16 \! \sum_{I < J < K} \fp^I \fp^J \fp^K \right) c_1 (J)^2 \\
 & - 8 \left(\frac{1}{n_1}+\frac{1}{n_2}\right) \sum_{I = 1}^4 ( 2 - \Delta_I ) \fp^I c_1 ( F_R ) c_1 ( J ) \, ,
\eea
and, with the same method used in section \ref{genm}, we can  extract from it the exact central charge of the CFT$_2$ and the levels of the various $\U(1)$ symmetries.
Defining
\be
 \Theta_{\text{KW}} ( \fp ) = \sum_{ \substack{I < J \\ (\neq K)}}^4 \fp^I \fp^J ( \fp^K )^4 - 2 \sum_{I , J = 1}^4 \fp^I \fp^J \prod_{K = 1}^4 \fp^K \, ,
\ee
we find the exact central charge
\be
 c^{\text{CFT}} = \frac{64}{9} a_{4\rd} \? \frac{( n_1 - n_2 )^2 ( \fp_1 \fp_2 \fp_3 + \fp_1 \fp_4 \fp_3 + \fp_2 \fp_4 \fp_3 + \fp_1 \fp_2 \fp_4 )}
 {( n_1^2 + n_1 n_2 + n_2^2 ) \prod_{I < J}^4 ( \fp^I + \fp^J ) + n_1 n_2 \? \Theta_{\text{KW}} ( \fp) } \, .
\ee
Furthermore, organizing the flavor symmetries in the basis $K_i= Q_i-Q_4$ for $i=1,2,3$ and $K_4=J$, we read the levels 
\bea
 k_{1,1} & = \frac{64}{27} a_{4 \rd} ( \fp^2 + \fp^3 )\, , \qquad k_{2,2} = \frac{64}{27} a_{4\rd} ( \fp^1 + \fp^3 ) \, , \qquad k_{3,3} = \frac{64}{27} a_{4\rd} ( \fp^1 + \fp^2 ) \, ,\\
 k_{4,4} & = - \frac{16}{27} a_{4\rd} \left( \frac{3}{16} \left( \frac{1}{n_1} + \frac{1}{n_2} \right) \left( \frac{1}{n_1^2} - \frac{1}{n_2^2} \right)
 + \fp^1 \fp^2 \fp^3 + \fp^1 \fp^4 \fp^3+\fp^2 \fp^4 \fp^3+\fp^1 \fp^2 \fp^4 \right) , \\
 k_{1,2} & = k_{1,3} = k_{2,3} = \frac{32}{27} \? a_{4 \rd} \? (\fp^1+\fp^2+\fp^3-\fp^4 ) \, , \\
 k_{4, j} & = \frac{8}{27} a_{4 \rd} \left(\frac{1}{n_1}+\frac{1}{n_2}\right) ( \fp^4-\fp^j ) \, , \quad \text{ for } \quad j = 1, 2, 3 \, .
\eea

Notice that in our discussion we allowed a generic gauge for the {\it flavor} symmetries parameterized by the quantities $r_0^i$. One can choose, for example, the value $r_0^i=2/d$ for all models.  The effect of a change of gauge for the flavor symmetries can be reabsorbed by a shift of $\Delta_i$ in $c_r(\epsilon, \Delta_i)$ and leads to the same exact central charge  $c^{\text{CFT}}$. It leads however to a redefinition of charges.
In particular, the charge $J$ associated with the internal isometry would be shifted by a linear combinations of the flavor charges. These redefinitions should be taken into account when computing the levels and writing the charged Cardy formula, but all physical conclusions are obviously unchanged.

\subsection{The universal spindles in theories with 8 and 16 supercharges}\label{UNIV}
There are two simple universal cases where we can test \eqref{A2d:gluingG} against supergravity predictions.
As already mentioned, the $stu$ model in five dimensions admits truncations to both minimal $\cN=2$ and $\cN=4$ gauged supergravites and the corresponding solutions   can be uplifted to any AdS$_5$ compactification  with eight and sixteen supercharges, respectively (charge quantization conditions and regularity of the uplift allowing). This calls for a universal field theory counting of microstates  in $\cN=1$ and $\cN=2$ SCFTs, in the spirit of \cite{Benini:2015bwz,Azzurli:2017kxo,Bobev:2017uzs,Hosseini:2020mut}.

Let us consider first the case with eight supercharges,  which is a straightforward generalization of the  universal static case discussed in \cite{Ferrero:2020laf}. The rotating black spindle of section \ref{subsec:rotspindle} when restricted to equal electric charges $q_1=q_2=q_3$ is a solution of minimal gauged supergravity and, as such, it can be embedded in all  string and M-theory compactifications with an AdS$_5$ vacuum and eight supercharges \cite{Gauntlett:2007ma}.  For all these solutions, the entropy takes the universal form
\be\label{UnivEntropy}
 S_{\text{BH}} \equiv \log \rho_{\text{susy}} (Q_0 , J) = 2 \pi  \sqrt{\frac{c^{\text{CFT}}}{6} \left( Q_0 - \frac{J^2}{2 k} \right)} \, ,
\ee
with
\be\label{univN=12}
 c^{\text{CFT}} = \frac{4 a_{4\rd}}{3} \frac{( n_2 - n_1 )^3}{n_1 n_2 (n_1^2+n_1 n_2+n_2^2 )} \, , \qquad k = - \frac{4 a_{4\rd}}{27} \left( \frac{1}{n_1^3}-\frac{1}{n_2^3} \right) ,
\ee
where $a_{4\rd}=\frac{\pi }{8  G^{(5)}_{\text{N}} g_{(5)}^3 }$ is the exact central charge of the dual $\cN=1$ SCFT. These solutions can be interpreted as universal compactifications 
on the spindle with angular momentum and magnetic charges aligned with the {\it exact} four-dimensional $R$-symmetry $\sum_{i=1}^d \bar \Delta_i  Q_i$ of  the $\cN=1$ SCFT \footnote{Notice that the exact $R$-charges $\bar \Delta_i$ should satisfy some conditions for the construction to work, in particular they should be at least rational. This restrict the class of compactifications that can be used.}
\be
 \label{UnivMag}
 \fp^i =  \frac 1 2 \left ( \frac{1}{n_1} - \frac{1}{n_2} \right) \bar \Delta_i \, , \quad \text{ for } \quad i = 1, 2, \ldots, d \, .
\ee
It is easy to see that the  anomaly polynomial of the two-dimensional CFT \eqref{A2d:gluingG} collapses to the anomaly polynomial \eqref{Aequal} of  $\cN=4$ SYM with $\Delta_i=2/3$,\footnote{The best way of doing this computation is to use the gauge $r_0^i=\bar\Delta_i$. Since there are no flavor charges and \eqref{A2d:gluingG} is extremized at $\Delta_i=\bar\Delta_i$, we can restrict to $\Delta_i=\bar\Delta_i$. Recall that  the exact four-dimensional $R$-symmetry $\bar \Delta_i $ is obtained by extremizing \eqref{gen-a} and $ a_{4\rd} = \frac{9}{32} \sum_{i,j,k = 1}^d c_{ijk} \bar \Delta_i \bar\Delta_j  \bar\Delta_k$.}
\bea\label{Aequal2}
 \cA_{2\rd} = \frac{2 a_{4\rd}}{27} & \left(12 \left( \frac{1}{n_1} - \frac{1}{n_2} \right) c_1 (F_R)^2 
  + \left( \frac{1}{n_1^3} - \frac{1}{n_2^3} \right) c_1 (J)^2 - 6 \left( \frac{1}{n_1^2} - \frac{1}{n_2^2} \right) c_1 (F_R) c_1 (J) \right) ,
\eea
 as already computed  in \cite[(30)]{Ferrero:2020laf}. Notice that this expression  only depends on $a_{4\rd}$. The universal form of \eqref{UnivEntropy} then follows from the universality of the anomaly polynomial and the charged Cardy formula.

Consider now the case with sixteen supercharges, in the spirit of \cite{Hosseini:2020mut}.
The five-dimensional uplift of the static spindle of section \ref{subsec:staticspindle}  when restricted to the magnetic charges $m^2=m^3$ are solutions of the half-maximal gauged supergravity in  five dimensions. Such solutions can be embedded in all AdS$_5$ type II or M-theory backgrounds with sixteen supercharges \cite{Gauntlett:2007sm,Cassani:2019vcl,Malek:2017njj}. For these solutions the exact central charge of the corresponding CFT$_2$ is given  
by
\be
 \label{cCFT:stuUniv}
 c^{\text{CFT}} = \frac{12 a_{4\rd} \? \fp^1 (\fp^2)^2}{\frac{1}{ n_1 n_2}+ 2\fp^1 \fp^2+(\fp^2)^2}\, ,
\ee
where $\fp^i= 2 g m^i$ and $\fp^1+2\fp^2=\frac{1}{n_1}-\frac{1}{n_2}$. The universality of this formula follows again from the universality of the four-dimensional trial central charge for $\cN=2$ SCFTs \cite{Shapere:2008zf}.
As discussed in \cite{Hosseini:2020mut}, for all the theories with a holographic dual  in the large $N$ limit we have
\be 
 a_{4\rd} (\Delta_i) = \frac{27 a_{4\rd}}{8} \Delta_1 \Delta_2^2 \, ,
\ee
where $\Delta_1$ and $\Delta_2$, with $\Delta_1+  2 \Delta_2 =2$, are associated with the $\U(1)_R$ and  the Cartan generator of the $\SU(2)_R$ symmetry, respectively.%
\footnote{In a perturbative theory the 4d trial $R$-symmetry is $R(\Delta)= \Delta_1 r_1/2+\Delta_2 r_2 $, where  $r_1$ assigns charge $2$ to the chiral field $\phi_I$ in the vector multiplet
and zero to the chiral pairs $q_a,\tilde q_a$ in a hypermultiplet and  $r_2$ assigns charge zero to $\phi_I$ and charge $1$ to $q_a,\tilde q_a$.
Compared with \cite{Hosseini:2020mut} we rescaled $\Delta_{2}^{\text{there}} = 2 \Delta_{2}^{\text{here}}$, for ease of comparison with $\cN=4$ SYM.}
It is then straightforward to check that the  anomaly polynomial of the two-dimensional CFT \eqref{A2d:gluingG} collapses to the anomaly polynomial \eqref{Aequal} of  $\cN=4$ SYM with $\Delta_2=\Delta_3$.
The expression \eqref{cCFT:stuUniv} then follows from the same computation as in section \ref{genm}.

\section{Discussion and outlook}
\label{sect:Discussion and outlook}

In this paper we constructed supersymmetric near-horizon solutions describing dyonic rotating spindle black strings that can be embedded in AdS$_5\times S^5$ and successfully matched the corresponding density  of states 
with the charged Cardy formula. Many questions are left open.

First of all, we were able to construct horizon geometries only. Given the successful field theory analysis based on $\cN=4$ SYM, we  expect these solutions to arise as supersymmetric domain walls  that interpolate between the near horizon region and AdS$_5$. It would be interesting to construct the full interpolating solution by generalizing the ansatz in \cite{Hristov:2019mqp}.
Similarly, it would be interesting to find spindle solutions in more general compactifications than AdS$_5\times S^5$. The case of universal solutions has been discussed in section \ref{UNIV}. More general solutions are more difficult to find. In order to use gauged supergravity,  we need the existence of a consistent truncation and deal with supergravity theories with hypermultiplets. Some recent examples in the context of AdS  black strings obtained from twisted compactifications can be found in \cite{Hosseini:2020vgl}.

Another important question concerns the field theory interpretation of the spindle compactifications. From the four-dimensional point of view, we are considering a SCFT on a singular manifold. It would be interesting to  understand whether this makes sense as a field theory in the presence of defects or it has an interpretation as the reduction of a more complicated quantum field theory on a higher-dimensional smooth manifold. We have seen that the anomaly polynomial of the low-energy CFT in two dimensions computed from purely four-dimensional data correctly reproduces the holographic results in the bulk. On the other hand,  the anomaly polynomial of the two-dimensional CFT should be also computable from a ten-dimensional point of view using the anomaly inflow method developed in  \cite{Bah:2020jas}.  This could not only put the computation on firmer ground but it could also shed some light on the physics of the system.  

Another natural question is whether the anomaly and other physical quantities, in particular the elliptic genus of the CFT$_2$, can be encoded in a higher-dimensional supersymmetric index, which could give information also at finite $N$. In the analogous case of  compactifications on the sphere, the relevant quantity  is the refined topologically twisted index \cite{Benini:2015noa}, which is the  partition function on $T^2\times S^2$ with a topological twist and an $\Omega$-background along $S^2$. This has been explicitly discussed in \cite{Hosseini:2016cyf,Hosseini:2019lkt}. Here the relevant index should be associated with a refined partition function on $T^2 \times \bW \bP^1_{[n_1, n_2]}$, with an appropriate prescription for dealing with the singularities. We notice that the supersymmetry on the spindle is not realized with a topological twist, but it is still supported by a magnetic flux \eqref{fluxSusy}. The gluing formulae  \eqref{gluing} and \eqref{A2d:gluingG} suggest that the relevant index for the spindle could be obtained by fusing two copies of the four-dimensional
holomorphic block \cite{Beem:2012mb} with fluxes $1/2n_1$ and $1/2n_2$ for the $R$-symmetry, respectively.

Finally, we notice that there exist also  supersymmetric accelerating and spinning black holes with conical singularities  in four dimensions. These can be interpreted as domain walls that interpolate between (conformal) AdS$_4$ and a warped product of AdS$_2$ and the spindle  and can be embedded in AdS$_4\times S^7$  \cite{Ferrero:2020twa}. There are many analogies with the situation considered in this paper. The near-horizon geometries can be found again with the formalism in \cite{Hristov:2019mqp}. Moreover, it is easy to check that an entropy functional for these black holes can be written by gluing two gravitational blocks with a formula similar to \eqref{gluing} using the prepotential  $\cF_{4\rd}=\sqrt{X^0 X^1 X^2 X^3}$. This corresponds to the familiar fact that, while the entropy of supersymmetric black objects in AdS$_5$ is controlled by anomalies and the central charge $a_{4\rd}(\Delta)$ of the dual four-dimensional SCFT, the entropy of black holes in AdS$_4$ is controlled by the free energy on $S^3$ of the dual three-dimensional SCFT, which for the ABJM theory \cite{Aharony:2008ug} reads $F_{S^3}=\frac{4 \pi \sqrt{2}\? k^{1/2} N^{3/2}}{3}\sqrt{\Delta_0 \Delta_1 \Delta_2 \Delta_3}$ in the large $N$ limit \cite{Jafferis:2011zi}. This again suggests the possible existence of an index obtained by fusing two copies of the three-dimensional
holomorphic block  with fluxes $1/2n_1$ and $1/2n_2$ for the $R$-symmetry. 

The present work, taken together with \cite{Ferrero:2020laf,Ferrero:2020twa}, also raises a more general question outside the realm of holography and microscopic entropy counting. We already noted that the solutions with conical singularities actually represent the most generic cases in lower-dimensional supergravity theories with a non-vanishing scalar potential, where no special restrictions are imposed on the black hole geometry and asymptotic charges. The fact that we can make sense of the supersymmetric solutions holographically might suggest the interesting possibility that the spindle horizons are also relevant for more realistic thermal black hole models. 

We leave all these questions for future work. 

\section*{Acknowledgments}
SMH is supported in part by WPI Initiative, MEXT, Japan at IPMU, the University of Tokyo, JSPS KAKENHI Grant-in-Aid (Wakate-A), No.17H04837 and JSPS KAKENHI Grant-in-Aid (Early-Career Scientists), No.20K14462.
KH is supported in part by the Bulgarian NSF grants DN08/3, N28/5, and KP-06-N 38/11.
AZ is partially supported by the INFN, the ERC-STG grant 637844-HBQFTNCER, and the MIUR-PRIN contract 2017CC72MK003.

\bibliographystyle{ytphys}
\bibliography{Spindle}

\end{document}